\RequirePackage{fix-cm}
\documentclass{svjour3}                     
\smartqed  
\usepackage{graphicx}

\usepackage{tikz}

\usepackage[english]{babel}
\usepackage{caption} 
\usepackage{mathptmx}      
\journalname{Finance and Stochastics}

\usepackage{amsmath,amsfonts}
\usepackage[normalem]{ulem}
\usepackage{amssymb}

\usepackage[hyphens]{url}
\usepackage{hyperref}
\usepackage{verbatim}
\usepackage{enumitem}
\usepackage{color}
\usetikzlibrary{decorations.fractals}
\usetikzlibrary{patterns,arrows,shapes,decorations.pathreplacing}
\usepackage{bm}
\usepackage{appendix}

\usepackage{soul}

\numberwithin{equation}{section}

\newtheorem{thm}{Theorem}
\newtheorem{lem}[thm]{Lemma}

\newtheorem{prop}[thm]{Proposition}
\newtheorem{cor}[thm]{Corollary}

\newtheorem{rem}[thm]{Remark}
\allowdisplaybreaks[1]

\numberwithin{thm}{section}

\newcommand {\E}{\mathbb{E}}
\newcommand {\Q}{\mathbb{Q}}
\renewcommand {\P}{\mathbb{P}}

\makeatletter
\newcommand {\@setcopyright}{}
\newcommand {\serieslogo@}{}
\makeatother

\begin{document}

\title{Optimal Consumption with Reference to Past Spending Maximum
}

\titlerunning{Optimal consumption with reference to past spending maximum}        

\author{Shuoqing DENG         \and
        Xun LI         \and
        Huy\^{e}n PHAM         \and
        Xiang YU
}


\institute{Shuoqing DENG \at
              Department of Mathematics, University of Michigan, Ann Arbor, USA. \\
              \email{shuoqing@umich.edu}           
           \and
           Xun LI \at
              Department of Applied Mathematics, The Hong Kong Polytechnic University, Kowloon, Hong Kong. \\
              \email{li.xun@polyu.edu.hk}  
           \and
           Huy\^{e}n PHAM \at
              LPSM, Universit\'e de Paris and CREST-ENSAE, Paris, France. \\
              \email{pham@lpsm.paris} 
           \and
           Xiang YU \at
              Department of Applied Mathematics, The Hong Kong Polytechnic University, Kowloon, Hong Kong.\\
              \email{xiang.yu@polyu.edu.hk} 
}

\date{Received: date / Accepted: date}

\maketitle

\begin{abstract}
This paper studies the infinite-horizon optimal consumption with a path-dependent reference under exponential utility. The performance is measured by the difference between the nonnegative consumption rate and a fraction of the historical consumption maximum. The consumption running maximum process is chosen as an auxiliary state process, and hence the value function depends on two state variables. The Hamilton-Jacobi-Bellman (HJB) equation can be heuristically expressed in a piecewise manner across different regions to take into account all constraints. By employing the dual transform and smooth-fit principle, some thresholds of the wealth variable are derived such that a classical solution to the HJB equation and the feedback optimal investment and consumption strategies can be obtained in closed form in each region. A complete proof of the verification theorem is provided, and numerical examples are presented to illustrate some financial implications.
\end{abstract}

\section{Introduction}\label{sec1}

The Merton problem, firstly studied in Merton \cite{Mert1} and \cite{Mert2}, has been one of the milestones in quantitative finance, which bridges investment decision making and some advanced mathematical tools such as PDE theories and stochastic analysis. By dynamic programming principle, one can solve the stochastic control problem by looking for the solution of the associated HJB equation. Isoelastic utility and exponential utility have attracted dominant attention in academic research as they enjoy the merits of homogeneity and scaling property. In abundant work on terminal wealth optimization, the value function can be conjectured in some separation forms or the change of variables can be applied. Consequently, the dimension reduction can be exercised to simplify the HJB equation. When the intermediate consumption is taken into account, the study of exponential utility becomes relatively rare in the literature due to its unnatural allowance of negative consumption behavior. To be precise, as the exponential utility is defined on the whole real line, the resulting optimal consumption by the first order condition can be negative in general. For technical convenience, some existing literature such as Merton \cite{Mert1}, Vayanos \cite{Vayanos}, Liu \cite{LiuH} and many subsequent work simply ignore the constraint or interpret the negative consumption by different financial meanings so that the non-negativity constraint on control can be avoided.

The case of exponential utility with non-negative consumption has been examined before in Cox and Huang \cite{CoxHuang} by using the martingale method, in which the optimal consumption can be expressed in an integral form using the state price density process. As shown in Cox and Huang \cite{CoxHuang}, the value function and the optimal consumption differ substantially from the case when the constraint is neglected. Some technical endeavors are required to fulfill the non-negativity constraint on the control process. In the present paper, we revisit this problem under the exponential utility binding with the non-negativity constraint on consumption rate. In addition, our study goes beyond the conventional time separable utilities and we aim to investigate the consumption behavior when an endogenous reference point is included inside the utility. Our proposed preference concerns how far the investor is away from the past consumption maximum level, and this intermediate gap is chosen as the metric to generate the utility of the investor in a dynamic way. Due to the consumption running maximum process in the utility, the martingale method developed in Cox and Huang \cite{CoxHuang} can no longer handle our path-dependent optimization problem because it is difficult to conjecture the correct dual processes and the associated dual problem.

Our research is mainly motivated by the psychological viewpoint that the consumer's satisfaction level and risk tolerance sometimes depend on recent changes instead of absolute rates. Some large amount of expenditures, such as purchasing a car, a house or some luxury goods, not only spur some long term continuing spending for maintenance and repair, but also lift up the investor's standard of living gradually. A striking decline in future consumption plan may result in intolerable disappointment and discomfort. To depict the quantitative influence of the relative change towards the investor's preference, it is reasonable to consider the utility that measures the distance between the consumption rate and a proportion of the past consumption peak. On the other hand, during some economic recession periods such as recent global economy battered by Covid-19, it is unrealistic to mandate that the investor needs to catch up with the past spending maximum all the time. To capture the possibility that the investor may strategically decrease the consumption budget to fall below the benchmark so that more wealth can be accumulated to meet the future higher consumption plan, we choose to work with the exponential utility that is defined on the positive real line. As a direct consequence, the investor can bear a negative gap between the current consumption and the reference level. The flexibility to compromise the consumption plan below the reference point from time to time makes the model suitable to accommodate more versatile market environments.

Utility maximization with a reference point has become an important topic in behavioral finance, see Tversky and Kahneman \cite{tvekah92},
He and Zhou \cite{He1}, He and Yang \cite{He2} and He and Strub \cite{He3} on portfolio management with either a fixed or an adaptive reference level. Our paper differs from the previous work as we do not distinguish the utility on gain and loss separately and our reference process is dynamically updated by the control itself. The impact of the path-dependent reference generated by the past consumption maximum becomes highly implicit in our model, which makes the mathematical problem appealing. Our formulation is also closely related to the consumption habit formation preference, which measures the deviation of the consumption from the standard of living conventionally defined as the weighted average of consumption integral. See some previous work on addictive consumption habit formation in Constantinides \cite{constantinides1990habit}, Detemple and Zapatero \cite{detemple1992optimal}, Schroder and Skiadas \cite{schroder2002isomorphism}, Munk \cite{munk2008portfolio}, Englezos and Karatzas \cite{englezos2009utility}, Yu \cite{yu2015utility}, \cite{yu2017} and non-addictive consumption habit formation in Detemple and Karatzas \cite{DepKart}. Recently, there are some emerging research on the combination of the reference and the habit formation, see Curatola \cite{Curatola17} and Bilsen et al. \cite{Bilsen17}, in which the reference level is generated by the endogenous habit formation process and different utility functions are equipped when the consumption is above and below the habit respectively. It will be an interesting future work to consider this S-shaped utility defined on the difference between the consumption and the consumption peak reference level and investigate the structure of the optimal consumption. Among the aforementioned work, it is worth noting that Detemple and Karatzas \cite{DepKart} considers the utility defined on the whole real line and also permits the admissible consumption to fall below the habit level from time to time. That is, the consumption habit is not addictive. Detemple and Karatzas \cite{DepKart} extends the martingale method in Cox and Huang \cite{CoxHuang} by using the adjusted state price density process, which produces a nice construction of the optimal consumption in the complete market model. However, the duality approach in Detemple and Karatzas \cite{DepKart} may not be applicable to our problem due to the presence of the running maximum process. 

One main contribution of the present paper is to show that the path-dependent control problem can be solved under the umbrella of dynamic programming and PDE approach. Comparing with the existing literature, the utility measures the difference between the control and its running maximum and the non-negativity constraint on consumption is imposed. The standard change of variables and the dimension reduction can not be applied, and we confront a value function depending on two state variables, namely the wealth variable $x\in \mathbb{R}_+$ and the reference level variable $h\in\mathbb{R}_+$. By noting that the consumption control is restricted between 0 and the peak level, we first heuristically derive the HJB equation in different forms based on the decomposition of the domain $\{(x,h)\in \mathbb{R}_+\times \mathbb{R}_+\}$ into disjoint regions of $(x,h)$ such that the feedback optimal consumption satisfies (i) $c^*(x,h)=0$; (ii) $0<c^*(x,h)<h$; (iii) $c^*(x,h)=h$. To overcome the obstacle from nonlinearity, we apply the dual transformation only with respect to the state variable $x$ and treat $h$ as the parameter that is involved in some free boundary conditions. The linearized dual PDE can be handled as a piecewise ODE problem with the parameter $h$. By using smooth-fit principle and some intrinsic boundary conditions, we obtain the explicit solution of the ODE that enables us to express the value function, the feedback optimal investment and consumption in terms of the primal variables after the inverse transform. We are able to find $x_{\text{zero}}(h)$, $x_{\text{modr}}(h)$, $x_{\text{aggr}}(h)$ and $x_{\text{lavs}}(h)$, indicating thresholds of zero consumption, moderate consumption, aggressive consumption and lavish consumption for the wealth variable $x$ as nonlinear functions of the variable $h$. The feedback optimal consumption can be characterized in the way that: (i) $c^*(x,h)=0$ when $x\leq x_{\text{zero}}(h)$; (ii) $0<c^*(x,h)<\lambda h$ when $x_{\text{zero}}(h)<x<x_{\text{modr}}(h)$; (iii) $\lambda h\leq c^*(x,h)<h$ when $x_{\text{modr}}(h)\leq x<x_{\text{aggr}}(h)$; (iv) $c^*(x,h)=h$ but the instant running maximum process $H_t^*$ remains flat when $x_{\text{aggr}}(h)\leq x<x_{\text{lavs}}(h)$; (v) $c^*(x,h)=h$ and the instant $c_t^*$ creates a new globall maximum level when $x=x_{\text{lavs}}(h)$. Moreover, due to the presence of the running maximum process inside the utility, the proof of the verification theorem involves many technical and non-standard arguments.

Building upon the closed-form value function and feedback optimal controls, some numerical examples are presented. The impacts of the variable $h$ and the reference degree parameter $\lambda$ on all boundary curves $x_{\text{zero}}(h)$, $x_{\text{modr}}(h)$, $x_{\text{aggr}}(h)$ and $x_{\text{lavs}}(h)$ can be numerically illustrated. We also perform sensitivity analysis of the value function, the optimal consumption and portfolio on some model parameters, namely the reference degree parameter, the mean return and the volatility of the risky asset, and discuss some quantitative properties and their financial implications.

The remainder of the paper is organized as follows. Section \ref{sec:introduc} introduces the market model and formulates the control problem under the utility with the reference to consumption peak. Section \ref{sec:solveHJB} presents the associated HJB equation for $0<\lambda<1$ and some heuristic results to derive its explicit solution. Some numerical examples are presented in Section \ref{sec:numerics}. Section \ref{sec:proofs} provides the proof of the verification theorem and other auxiliary results in the previous sections. At last, the main result of the extreme case $\lambda=1$ is given in Appendix \ref{appA} and the lengthy proof of one auxiliary lemma is given in Appendix \ref{appB}.

\section{Market Model and Problem Formulation}\label{sec:introduc}
Let $(\Omega, \mathcal{F}, \mathbb{F}, \mathbb{P})$ be a filtered probability space, in which $\mathbb{F}=(\mathcal{F}_t)_{t\geq 0}$ satisfies the usual conditions. We consider a financial market consisting of one riskless asset and one risky asset. The riskless asset price satisfies $dB_t=rB_tdt$ where $r\geq 0$ represents the constant interest rate. The risky asset price follows the dynamics
\begin{equation}
dS_t=S_t\mu dt+S_t\sigma dW_t,\nonumber
\end{equation}
where $W$ is an $\mathbb{F}$-adapted Brownian motion and the drift $\mu$ and volatility $\sigma>0$ are given constants. The Sharpe ratio parameter is denoted by $\kappa:= \frac{\mu-r}{\sigma}$. It is worth noting that our mathematical arguments and all conclusions can be readily generalized to the model with multiple risky assets as long as the market is complete. For the sake of simple presentation, we shall only focus on the model with a single risky asset. It is assumed that $\kappa>0$ from this point onwards, i.e. $\mu>r$ that the return of the risky asset is higher than the interest rate.

Let $(\pi_t)_{t\geq 0}$ represent the dynamic amount that the investor allocates in the risky asset and $(c_t)_{t\geq 0}$ denote the dynamic consumption rate of the investor. The resulting self-financing wealth process $(X_t)_{t\geq 0}$ satisfies
\begin{equation}\label{originw}
dX_t=rX_tdt+ \pi_t(\mu -r)dt+\pi_t\sigma dW_t-c_tdt,\ \ \ t\geq 0,
\end{equation}
with the initial wealth $X_0=x\geq 0$.

The consumption-portfolio pair $(c, \pi)$ is said to be \textit{admissible}, denoted by $(c, \pi)$ $\in$ $\mathcal{A}(x)$, if the consumption rate $c_t\geq 0$ a.s. for all $t\geq 0$, $c$ is $\mathbb{F}$-predictable, $\pi$ is $\mathbb{F}$-progressively measurable and both satisfy the integrability condition $\int_0^{\infty} (c_t+\pi_t^2)dt<\infty$ a.s. Moreover, no bankruptcy is allowed in the sense that $X_t\geq 0$ a.s. for $t\geq 0$.

Let us focus on the exponential utility $U(x)=-\frac{1}{\beta}e^{-\beta x}$ in the present paper with $\beta>0$, $x$ $\in$ $\mathbb{R}$.
We are interested in the following infinite horizon utility maximization defined on the difference between the current consumption rate and its historical running maximum that
\begin{align}\label{primalvalue}
u(x, h)=\sup_{(\pi, c)\in\mathcal{A}(x)}\mathbb{E}\left[\int_0^{\infty} e^{-\rho t}U(c_t-\lambda H_t)dt\right],
\end{align}
where 
\begin{equation}
H_t=\max{\{h,\ \ \sup_{s\leq t} c_s\}},\ \ H_0=h\geq 0,\nonumber
\end{equation}
and the proportional constant $0\leq \lambda\leq 1$ depicts the intensity towards the reference level that the investor adheres to the past spending pattern. Here, $H_0=h\geq 0$ describes the reference level of the consumption that the individual aims to surpass at the initial time.

One advantage of the exponential utility resides in the flexibility that the optimal consumption $c^*$ can fall below the reference level $\lambda H^*$, which matches better with the real life situation that the investor can bear some unfulfilling consumption during the economic recession periods. That is, to achieve the value function, it is not necessary for the optimal consumption control to exceed the reference level at any time. Meanwhile, the non-negativity constraint $c_t\geq 0$ a.s. should be actively enforced for all time $t\geq 0$. This control constraint spurs some new challenges when we handle the associated HJB equation using dynamic programming arguments in subsequent sections. We shall only focus on the more interesting case $0<\lambda<1$ in the main body of this paper. The extreme case $\lambda=0$ is a standard Merton problem under exponential utility, which will be omitted. Some main results in the other extreme case $\lambda=1$ are reported in Appendix \ref{appA}.

\section{Main Results} \label{sec:solveHJB}

For ease of presentation and technical convenience, we only consider the case that $\rho=r>0$. The general cases $(i)$ $\rho\neq r>0$ and $(ii)$ $r=0$ and $\rho\geq 0$ can be handled similarly, leading to more complicated formulas. Additional parameter assumptions are therefore required in these general cases to support the optimality in the verification proof, which are beyond the scope of this paper. To embed the control problem into a Markovian framework and derive the HJB equation using dynamic programming arguments, we treat both $X_t$ and $H_t$ as the controlled state processes given the control policy $(c,\pi)$. The value function $u(x,h)$ depends on both variables $x\geq 0$ and $h\geq 0$, namely the initial wealth and the initial reference level. Let us consider
\begin{equation*}
\Gamma_t:=e^{-r t} u(X_t, H_t)+\int_0^t e^{-r s}U(c_s-\lambda H_s)ds.
\end{equation*}
Heuristically, by the martingale optimality principle, we have that $(\Gamma_t)_{t\geq 0}$ is a local supermartingale under all admissible controls and $(\Gamma_t)_{t\geq 0}$ is a local martingale under the optimal control (if it exists). If the function $u(x,h)$ is smooth enough, by applying It\^{o}'s formula to the process $(\Gamma_t)_{t\geq 0}$, we can derive that
\begin{align*}
e^{r t}d\Gamma_t=&\left( -r u+u_x\big(rX_t+\pi_t(\mu-r)-c_t\big)+\frac{1}{2}\sigma^2\pi_t^2 u_{xx}+U(c_t- \lambda H_t) \right)dt\\
&+u_hdH_t+u_x\pi_t \sigma dW_t,
\end{align*}
which heuristically leads to the associated HJB variational inequality
\begin{equation}\label{HJB_eqn}
\left\{
\begin{array}{rcl}
\underset{c\in [0,h], \pi\in\mathbb{R}}{\sup}\left( -r u+u_x\big(rx+\pi(\mu-r)-c\big)+\frac{1}{2}\sigma^2\pi^2 u_{xx}-\frac{1}{\beta}e^{\beta(\lambda h-c)}\right) &=& 0,  \\
u_h(x,h)  & \leq & 0,
\end{array}
\right.
\end{equation}
for $x\geq 0$, $h \geq 0$. The local martingale property of $u(X^*_t, H^*_t)$ under the optimal control ($c^*$, $\pi^*$) requires that $u_h(X^*_t, H^*_t)=0$ whenever the process $H_t^*$ strictly increases, i.e., the current consumption rate $c_t^*$ creates the new historical maximum level that $H_t^*=c_t^*$ and $c_t^*>H_s^*$ for $s<t$. This motivates us to mandate an important free boundary condition that $u_h(x,h)=0$ on some set of $(x,h)$ that will be determined explicitly later in \eqref{martingale_condition} in the section when we analyze the associated HJB equation.

In the present paper, we aim to find some deterministic functions $\pi^{\ast}(x,h)$ and $c^{\ast}(x,h)$ to provide the feedback form of the optimal portfolio and consumption strategy. To this end, if $u(x,\cdot)$ is $C^2$ w.r.t the variable $x$, the first order condition gives the optimal portfolio in a feedback form by $\pi^{\ast}(x,h)=-\frac{\mu-r}{\sigma^2}\frac{u_x}{u_{xx}}$. The previous HJB variational inequality \eqref{HJB_eqn} can first be written as
\begin{equation} \label{ODE}
\sup_{c\in [0,h]} \left( -\frac{1}{\beta}e^{\beta(\lambda h-c)} - c u_x \right) -r u + rx u_x -\frac{\kappa^2}{2}\frac{u_x^2}{u_{xx}} = 0,\ \ \text{and}\ \ u_h\leq 0,\ \ \forall x\geq0,h\geq 0,
\end{equation}
together with the free boundary condition $u_h=0$ on some set of $(x,h)\in\mathbb{R}_+\times\mathbb{R}_+$ that will be characterized later.

\subsection{Heuristic solution to the HJB equation} \label{subsec:Heuristic}
In view that $0\leq c_t\leq H_t$, we first need to decompose the domain $(x,h)\in\mathbb{R}_+\times \mathbb{R}_+$ into three different regions such that the feedback optimal consumption strategy satisfies: (1) $c^*(x,h)=0$; (2) $0<c^*(x,h)<h$; (3) $c^*(x,h)=h$. By applying the first order condition to the HJB equation \eqref{ODE}, let us consider the auxiliary control $\hat c(x,h):= - \frac{1}{\beta}\ln u_x + \lambda h$, which facilitates the separation of the following regions:

\vspace{3mm}

\noindent \textit{Region I}: on the set $\mathcal{R}_1:=\{(x,h)\in\mathbb{R}_+\times \mathbb{R}_+:u_x(x,h)\geq e^{\lambda \beta h} \}$, we have $\hat c(x,h) \leq 0$, and the optimal consumption is therefore $c^*(x,h)=0$. The HJB variational inequality becomes
\begin{equation}\label{veq-1}
-\frac{1}{\beta}e^{\lambda \beta h} -r u+ rx u_x - \frac{\kappa^2 u_x^2}{2 u_{xx}} = 0, \ \text{and}\ \ u_h \leq 0.
\end{equation}
\textit{Region II}: on the set $\mathcal{R}_2:=\{(x,h)\in\mathbb{R}_+\times \mathbb{R}_+:e^{-(1-\lambda)\beta h} < u_x(x,h)<  e^{\lambda \beta h} \}$, we have $0 <  \hat c(x,h) <  h$, and the optimal consumption is therefore $c^* = -\frac{1}{\beta} \ln u_x + \lambda h$. The HJB variational inequality becomes
\begin{equation}\label{veq-2}
-\frac{1}{\beta}u_x+ u_x (\frac{1}{\beta}\ln u_x - \lambda h) -r u + rx u_x -\frac{\kappa^2 u_x^2}{2 u_{xx}} = 0,\ \text{and}\ \  u_h \leq 0.
\end{equation}

\begin{rem}\label{comparec-lambdah}
Based on $c^* = -\frac{1}{\beta}\ln u_x+ \lambda h$ in Region II, we know that $c^*<\lambda H^*$ if and only if $(x,h)$ is in the subset $\left\{(x,h)\in\mathbb{R}_+\times \mathbb{R}_+:1<u_x(x,h)< e^{\lambda \beta h}\right\}$. This subset can be further expressed later in Remark \ref{discussthresh} as a threshold (depending on $h$) of the wealth level $x$.
\end{rem}

\textit{Region III}: on the set $\mathcal{R}_3:=\{(x,h)\in\mathbb{R}_+\times \mathbb{R}_+:u_x(x,h) \leq  e^{-(1-\lambda)\beta h} \}$, we have that $\hat c(x,h) \geq h$ and the optimal consumption is $c^*(x,h)= h$, which indicates that the instant consumption rate $c_t^*$ coincides with the running maximum process $H_t^*$. However, two subtle cases may occur that motivate us to split this region further:
\begin{itemize}
\item[(i)] In a certain region (to be determined), the historical maximum level is already attained at some previous time $s$ before time $t$ and the current optimal consumption rate is either to revisit this maximum level from below or to sit on the same maximum level. This is the case that the running maximum process $H_t$ keeps flat from time $s$ to time $t$, and the feedback consumption takes the form $c_t^*=H_s^*$ for some $s<t$. 
\item [(ii)] In the complementary region, the optimal consumption rate creates a new record of the maximum level that is strictly larger than its past consumption, and the running maximum process $H_t$ is strictly increasing at the instant time $t$. This corresponds to the case that $c_t^*=H_t^*$ is a singular control and $c_t^*>H_s^*$ for $s<t$ and we have to mandate the free boundary condition $u_h(x,h)=0$ from the martingale optimality condition. 
\end{itemize}

Restricted on the set $\{(x,h)\in\mathbb{R}_+\times \mathbb{R}_+:u_x(x,h)\leq  e^{-(1-\lambda)\beta h}  \}$, the case $(ii)$ suggests us to treat the $H_t^*=c^*_t$ as a singular control instead of the state process. That is, the dimension of the problem can be reduced and we can first substitute $h=c$ in \eqref{ODE} and then apply the first order condition to $-\frac{1}{\beta}e^{\beta(\lambda c - c)} - c u_x$ with respect to $c$. We can obtain the auxiliary control $\hat{c}(x):=\frac{1}{\beta(\lambda-1)} \ln (\frac{u_x}{1-\lambda})$. It is then convenient to see that $c_t^*$ can update $H_t^*$ to a new level if and only if the feedback control $c_t^*=\hat{c}(X_t^*)\geq H_t^*$ so that $H_t^*$ is instantly increasing. We therefore separate Region III into three subsets:  \\
\ \\
\textit{Region III-(i)}: on the set $\mathcal{D}_1:=\{(x,h)\in\mathbb{R}_+\times \mathbb{R}_+: (1-\lambda) e^{-(1-\lambda)\beta h}<u_x(x,h) \leq e^{-(1-\lambda)\beta h} \}$, we have a contradiction that $\hat{c}(x)<h$, and therefore $c^*_t$ is not a singular control. We should follow the previous feedback form $c^*(x,h) = h$, in which $h$ is a previously attained maximum level. The corresponding running maximum process remains flat at the instant time. In this region of $(x,h)$, we only know that $u_h(x,h)\leq 0$ as we have $dH_t=0$. The HJB variational inequality is written by
\begin{equation}\label{veq-3}
 -\frac{1}{\beta}e^{\beta(\lambda h-h)} - h u_x -r u + rx u_x -\frac{\kappa^2 u_x^2}{2 u_{xx}} = 0,\ \text{and}\ \ u_h \leq 0.
\end{equation}
\ \\
\textit{Region III-(ii)}: on the set $\mathcal{D}_2:=\{(x,h)\in\mathbb{R}_+\times \mathbb{R}_+: u_x(x,h)=(1-\lambda) e^{-(1-\lambda)\beta h}\}$, we get $\hat{c}(x)=h$ and the feedback optimal consumption is $c^*(x,h)=\frac{1}{\beta(\lambda-1)} \ln (\frac{u_x}{1-\lambda})=h$. This corresponds to the singular control $c_t^*$ that creates a new peak for the whole path and $H_t^*=c_t^*=\frac{1}{\beta(\lambda-1)} \ln (\frac{u_x(X_t^*,H_t^*)}{1-\lambda})$ is strictly increasing at the instant time so that $H_t^*>H_s^*$ for $s<t$ and we must require the following free boundary condition that
\begin{equation}\label{martingale_condition}
u_h(x,h)=0\ \ \text{on}\ \left\{(x,h)\in\mathbb{R}_+\times \mathbb{R}_+: u_x(x,h)=(1-\lambda) e^{-(1-\lambda)\beta h}\right\}.
\end{equation}
In this region, the HJB equation follows the same PDE \eqref{veq-3} but together with the free boundary condition \eqref{martingale_condition}.\\
\ \\
\textit{Region III-(iii)}: on the set $\mathcal{D}_3:=\{(x,h)\in\mathbb{R}_+\times \mathbb{R}_+: u_x(x,h)<(1-\lambda) e^{-(1-\lambda)\beta h}\}$, we get $\hat{c}(x)>h$. This indicates that the initial reference level $h$ is below the feedback control $\hat{c}(x)$, and the optimal consumption is again $c^*(x,h)=\frac{1}{\beta(\lambda-1)} \ln (\frac{u_x}{1-\lambda})$. As the running maximum process $H_t^*$ is updated immediately by $c_t^*$, the feedback optimal consumption pulls the associated $H_{t-}^*$ upward to the new value $\frac{1}{\beta(\lambda-1)} \ln (\frac{u_x(X_t^*, H_t^*)} {1-\lambda})$ in the direction of $h$ while $X_t^*$ remains the same, in which $u(x,h)$ is the solution of the HJB equation \eqref{veq-3} on the set $\mathcal{D}_2$. This suggests that for any given initial value $(x,h)$ in the set $\mathcal{D}_3$, the feedback control $c^*(x,h)$ pushes the value function jumping immediately to the point $(x, \hat{h})$ on the set $\mathcal{D}_2$ where $\hat{h}=\frac{1}{\beta(\lambda-1)} \ln (\frac{u_x(x,\hat{h})}{1-\lambda})$ for the given value of $x$.

In summary, it is sufficient for us to only concentrate $(x,h)$ on the \textit{effective domain} of the stochastic control problem that
\begin{align}\label{primedomain}
\mathcal{C}:=\left\{(x,h)\in\mathbb{R}_+\times \mathbb{R}_+: u_x(x,h)\geq (1-\lambda) e^{-(1-\lambda)\beta h}\right\}.
\end{align}
Equivalently, we have $\mathcal{C}=\mathcal{R}_1\cup\mathcal{R}_2\cup\mathcal{D}_1\cup\mathcal{D}_2\subset \mathbb{R}_+^2$. Notice that, the only possibility for $(x,h)\in \mathcal{D}_3=\mathcal{C}^c$ occurs at the initial time $t=0$, and the value function is just equivalent to the value function of $(x,\hat{h})$ on the boundary $\mathcal{D}_2$ with the same $x$. In other words, if the controlled process $(X_0^*, H_0^*)$ starts from $(x,h)$ in the region $\mathcal{C}$, then $(X_t^*, H_t^*)$ will always stay inside the region $\mathcal{C}$. On the other hand, if the process $(X_0^*, H_0^*)$ starts from the value $(x,h)$ inside the region $\mathcal{D}_3$, the optimal control enforces an instant jump (and the only jump) of the process $H^*$ from $H_{0-}^*=h$ to $H_0^*=\hat{h}$ on the set $\mathcal{D}_2$, and both processes $X_t^*$ and $H_t^*$ are continuous processes diffusing inside the effective domain $\mathcal{C}$ afterwards for $t>0$.

On the other hand, observe that as the wealth level $x$ declines to zero, the consumption rate $c$ will reach zero at some $x^*$(to be determined). If $x$ continues to decrease to $0$, the optimal investment $\pi$ should also drop to $0$. Otherwise, we will confront the risk of bankruptcy by keeping trading with the nearly $0$ wealth. Using the optimal portfolio $\pi^{\ast}(x,h)=-\frac{\mu-r}{\sigma^2}\frac{u_x}{u_{xx}}$, the boundary condition can be described by
\begin{align}\label{bound0-1}
\lim_{x\rightarrow 0}\frac{u_x(x,h)}{u_{xx}(x,h)}=0.
\end{align}
In addition, if we start with $0$ initial wealth, the wealth level will never change as there is no trading according to the previous condition, and the consumption should stay at $0$ consequently. That is, we have another boundary condition that
\begin{align}\label{bound0-2}
\lim_{x\rightarrow 0}u(x,h)= \int_0^{+\infty} - \frac{1}{\beta}e^0 e^{-r t} dt =-\frac{1}{r\beta}.
\end{align}
On the other hand, as the wealth tends to infinitely large, one can consume as much as possible and a small variation in the wealth has a negligible effect on the change of the value function. It thus follows that
\begin{align}\label{boundinf}
\lim_{x\rightarrow +\infty}u(x,h)=0\ \ \text{and}\ \ \lim_{x\rightarrow +\infty}u_x(x,h)= 0.
\end{align}

To ensure the global regularity of the solution, we also need to impose the smooth-fit conditions along two free boundaries of $(x,h)$ such that $u_x(x,h)= e^{\lambda \beta h}$, $u_x(x,h) = e^{-(1-\lambda)\beta h}$, which separate the regions as discussed above.

We can then employ the dual transform approach to linearize the HJB equation. In particular, we apply the dual transform only with respect to the variable $x$ and treat the variable $h$ as a parameter. That is, for each fixed $h\geq 0$, we consider $x\geq 0$ such that $(x,h)\in\mathcal{C}$ and define the dual function on the domain $y\geq  (1-\lambda) e^{-(1-\lambda)\beta h} $ that
\begin{align*}
v(y,h) & := \sup_{\substack{(x,h)\in\mathcal{C},\\ x\geq 0}} \big( u(x,h) - xy\big),\ \ y\geq (1-\lambda) e^{-(1-\lambda)\beta h} .
\end{align*}

For the given $(x,h)$, let us define $\hat{y}(x,h):=u_x(x,h)$ (short as $\hat{y}$), the dual representation implies $u(x,h)=v(\hat{y}, h)+x\hat{y}$ as well as $v_y(\hat{y}, h)=-x$. We then have
\begin{align*}
u_h(x,h)=\frac{\partial}{\partial h}(v(\hat{y},h)+x\hat{y})=v_h(\hat{y},h)+(v_y(\hat{y},h)+x)\frac{d\hat{y}}{dh}=v_h(\hat{y},h).
\end{align*}
In view of \eqref{martingale_condition}, we obtain the free boundary condition that
\begin{align}\label{freedul}
v_h(y,h)=0\ \ \text{on the set}\  \left\{(y,h)\in(0,+\infty)\times \mathbb{R}_+: y=(1-\lambda)e^{(\lambda-1)\beta h}\right\}.
\end{align}

To align with nonlinear HJB variational inequality \eqref{veq-1}, \eqref{veq-2}, \eqref{veq-3} in three different regions, the transformed dual variational inequality can be written as
\begin{equation} \label{EDPtildeU2}
 \frac{\kappa^2}{2} y^2 v_{yy} - r v =\left\{
\begin{aligned}
& \frac{1}{\beta}e^{\lambda \beta h}, & & \mbox{if } y \geq e^{\lambda \beta h}, \\
& \frac{1}{\beta}y - y \left(\frac{1}{\beta} \ln y - \lambda h\right),  & & \mbox{if } e^{(\lambda-1)\beta h} < y < e^{\lambda \beta h}, \\
& \frac{1}{\beta}e^{(\lambda-1)\beta h} + hy,  & & \mbox{if } (1-\lambda)e^{(\lambda-1)\beta h} \leq y \leq e^{(\lambda-1)\beta h},
\end{aligned}
\right.
\end{equation}
together with the free boundary condition \eqref{freedul}. As $h$ is regarded as a parameter, we shall fix $h$ and study the above equation as an ODE problem of the variable $y$.

By virtue of the duality representation, the boundary conditions in \eqref{boundinf} become
\begin{align}\label{yhcond-1}
\lim_{y\rightarrow 0}v_y(y,h)= -\infty\ \ \text{and}\ \ \lim_{y \rightarrow 0} (v(y,h)-yv_y(y,h))=0,
\end{align}
and the boundary conditions \eqref{bound0-1} and \eqref{bound0-2} at $x=0$ can be written as
\begin{align}\label{yhcond-2}
yv_{yy}(y,h) \rightarrow 0\ \text{and}\ \ v(y,h)-yv_y(y,h) \rightarrow -\frac{1}{r\beta}e^{\lambda \beta h}\ \ \text{as}\ \  v_y(y,h) \rightarrow 0.
\end{align}

Based on these boundary conditions, we can solve the dual ODE \eqref{EDPtildeU2} fully explicitly and its proof is given in Section \ref{otherproofs}.
\begin{prop}\label{dual_value_function}
Let $h\geq 0$ be a fixed parameter. Under the boundary conditions in \eqref{yhcond-1} and \eqref{yhcond-2} and the free boundary condition \eqref{freedul} as well as the smooth-fit conditions with respect to $y$ at boundary points $y=e^{\lambda\beta h }$ and $y=e^{(\lambda-1)\beta h}$, the ODE \eqref{EDPtildeU2} in the domain $y\geq (1-\lambda)e^{(\lambda-1)\beta h}$ admits the unique solution that
\begin{align*}
&v(y,h) =\\&\left\{
\begin{aligned}
& C_2(h) y^{r_2} - \frac{1}{r\beta} e^{\lambda \beta h}, & & \mbox{if } y \geq e^{\lambda \beta h}, \\
& C_3(h) y^{r_1} + C_4(h) y^{r_2}-\frac{y}{r\beta}+\frac{y}{r\beta } \left(\ln y-\lambda\beta h+\frac{\kappa^2}{2r}\right), & &\mbox{if } e^{(\lambda-1)\beta h} < y < e^{\lambda \beta h}, \\
& C_5(h) y^{r_1} + C_6(h) y^{r_2} -\frac{1}{r}hy-\frac{1}{r\beta} e^{(\lambda-1)\beta h}, & &\mbox{if } (1-\lambda)e^{(\lambda-1)\beta h} \leq y \leq e^{(\lambda-1)\beta h},
\end{aligned}
\right.
\end{align*}
where functions $C_2(h)$, $C_3(h)$, $C_4(h)$, $C_5(h)$ and $C_6(h)$ are given explicitly in \eqref{C2h}, \eqref{C3h}, \eqref{C4h}, \eqref{C5h} and \eqref{C6h} respectively that
\begin{align}
C_2(h) := & \frac{(1-\lambda)^{r_1-r_2}}{(r_1-r_2)\beta r} \left( \frac{1}{1-r_2} e^{(\lambda-1)(1-r_2)\beta h} -\frac{\lambda}{\lambda(1-r_2)-(r_1-r_2)} e^{\left( \lambda(1-r_2)-(r_1-r_2) \right)\beta h} \right)  \notag\\
&+ \frac{(1-r_1)\kappa^2}{2(r_1-r_2)\beta r^2}\left( e^{(\lambda-1)(1-r_2)\beta h}-e^{\lambda(1-r_2)\beta h} \right);\label{C2h}
\end{align}
\begin{align}
C_3(h):=\frac{(r_2-1) \kappa^2}{2(r_1-r_2)\beta r^2}e^{\lambda(1-r_1)\beta h}\label{C3h};
\end{align}
\begin{align}
C_4(h) := & \frac{(1-\lambda)^{r_1-r_2}}{(r_1-r_2)\beta r} \left( \frac{1}{1-r_2} e^{(\lambda-1)(1-r_2)\beta h} -\frac{\lambda}{\lambda(1-r_2)-(r_1-r_2)} e^{\left( \lambda(1-r_2)-(r_1-r_2) \right)\beta h} \right) \notag\\
&+ \frac{(1-r_1)\kappa^2}{2(r_1-r_2)\beta r^2}e^{(\lambda-1)(1-r_2)\beta h}\label{C4h};
\end{align}
\begin{align}
C_5(h) := \frac{(1-r_2) \kappa^2}{2(r_1-r_2)\beta r^2} \left(e^{(\lambda-1)(1-r_1)\beta h}- e^{\lambda (1-r_1)\beta h}\right);\label{C5h}
\end{align}
\begin{align}
C_6(h):= & \frac{(1-\lambda)^{r_1-r_2}}{(r_1-r_2)\beta r} \left( \frac{1}{1-r_2} e^{(\lambda-1)(1-r_2)\beta h} -\frac{\lambda}{\lambda(1-r_2)-(r_1-r_2)} e^{\left( \lambda(1-r_2)-(r_1-r_2) \right)\beta h} \right).\label{C6h}
\end{align}
Here, the constants $r_1>1$ and $r_2<0$ are two roots of the quadratic equation
$z^2 - z - \frac{2r}{\kappa^2} =0$, which are given by
\begin{align*}
r_{1,2}=\frac{1}{2} \Big( 1 \pm \sqrt{1+\frac{8 r}{\kappa^2}} ~ \Big).
\end{align*}

\end{prop}


We can now present the main result of this paper, which provides the optimal investment and consumption strategies in the piecewise feedback form using variables $y$ and $h$. The complete proof is deferred to Section \ref{sec:verification}.
\begin{thm}[Verification Theorem]\label{verthm}
Let $(x,h)\in\mathcal{C}$, where $\mathcal{C}$ is the effective domain \eqref{primedomain}. For $(y,h)\in (0,+\infty)\times [0,+\infty)$, let us define the feedback functions that
\begin{align}\label{feedbackcp}
c^{\dagger}(y,h) =\left\{
\begin{aligned}
&0, & & \mbox{if } y \geq e^{\lambda \beta h}, \\
&- \frac{1}{\beta}\ln y + \lambda h,  & & \mbox{if } e^{(\lambda-1)\beta h} < y < e^{\lambda \beta h}, \\
&h,  & & \mbox{if } (1-\lambda)e^{(\lambda-1)\beta h} <y \leq e^{(\lambda-1)\beta h}, \\
&\frac{1}{(\lambda-1)\beta} \ln \Big(\frac{1}{1-\lambda} y\Big),     & & \mbox{if } y = (1-\lambda)e^{(\lambda-1)\beta h},
\end{aligned}
\right.
\end{align}

\begin{align}\label{feedbackpi}
\begin{aligned}
&\pi^{\dagger}(y,h)= \frac{\mu-r}{\sigma^2}\left\{
\begin{aligned}
& \frac{2r}{\kappa^2}C_2(h) y^{r_2-1},  & & \mbox{if } y \geq e^{\lambda \beta h}, \\
& \frac{2r}{\kappa^2}C_3(h) y^{r_1-1} + \frac{2r}{\kappa^2}C_4(h) y^{r_2-1}+ \frac{1}{r\beta},  & & \mbox{if } e^{(\lambda-1)\beta h} < y < e^{\lambda \beta h}, \\
& \frac{2r}{\kappa^2}C_5(h) y^{r_1-1} + \frac{2r}{\kappa^2}C_6(h) y^{r_2-1}, & & \mbox{if } (1-\lambda)e^{(\lambda-1)\beta h} \leq y \leq e^{(\lambda-1)\beta h}.
\end{aligned}
\right.
\end{aligned}
\end{align}
We consider the process $Y_t(y):=ye^{r t} M_t$, where $M_t:= e^{-(r+\frac{\kappa^2}{2})t - \kappa W_t}$ is the discounted state price density process. Let the constant $y^*=y^*(x,h)$ be the unique solution to the budget constraint equation $\E [\int_0^{\infty}  c^{\dagger}(Y_t(y),H_t^{\dagger}(y)) M_t dt ]=x$, where 
$$
H_t^{\dagger}(y): =h\lor \sup_{s \leq t} c^{\dagger}(Y_s(y),H_s^{\dagger}(y)) = h \lor \Bigg( \frac{1}{(\lambda-1)\beta} \ln\left(\frac{1}{1-\lambda} \inf_{s \leq t} Y_s(y) \right) \Bigg)
$$
is the optimal reference process corresponding to any fixed $y >0$.
The value function $u(x,h)$ can be attained by employing the optimal consumption and portfolio strategies in the feedback form that $c_t^*=c^{\dagger}(Y^*_t,H_t^*)$ and $\pi_t^{*}=\pi^{\dagger}(Y^*_t,H_t^*)$, for all $t\geq 0$, where $Y_t^*:=Y_t(y^*)$ and $H_t^*=H_t^{\dagger}(y^*)$. 

The process $H_t^*$ is strictly increasing if and only if $Y^*_t=(1-\lambda)e^{(\lambda-1)\beta H_t^*}$. If we have $y^*(x,h)<(1-\lambda)e^{(\lambda-1)\beta h}$ at the initial time, the optimal consumption creates a new peak and brings $H_{0-}^*=h$ jumping immediately to a higher level $H_0^*=\frac{1}{(\lambda-1)\beta} \ln (\frac{1}{1-\lambda} y^*(x,h))$ such that $t=0$ becomes the only jump time of $H^*_t$.
\end{thm}

\begin{rem}
Note that the feedback optimal consumption $c^*_t=c^{\dagger}(Y_t^*, H_t^*)$ in \eqref{feedbackcp} is predictable. Indeed, if $Y^*_t>(1-\lambda)e^{(\lambda-1)\beta H^*_{t}}$, the optimal consumption at time $t$ is determined by the continuous process $Y^*_t$ and the past consumption maximum right before $t$, i.e. $H^*_{t-}$, which is predictable. In this case, the current consumption does not create the new maximum level. When $Y^*_t=(1-\lambda)e^{(\lambda-1)\beta H^*_{t}}$, the optimal consumption is determined directly by the continuous process $Y^*_t$, which is again predictable.
\end{rem}

In Theorem \ref{verthm}, the feedback controls are given in terms of the dual value function and the dual variables. In what follows, we show that the inverse transformation can be exercised so that the primal value function $u(x,h)$ and the feedback controls can be expressed by $x$ and $h$. In the proof of Theorem \ref{verthm}, we will take full advantage of the simplicity in the dual feedback controls and verify their optimality using the duality relationship and some estimations based on the dual process $Y^*_t=y^*e^{r t} M_t$. However, in the last step, to show the existence of a unique strong solution of the SDE \eqref{originw} under the optimal controls, we have to express the feedback controls in terms of $X_t^*$ and $H_t^*$ and therefore the inverse dual transform becomes necessary, which will be carefully established as follows.

By using the dual relationship that $v(y,h)=\sup_{x>0}(u(x,h)-xy)$, we have that the optimal choice $x$ satisfying $u_x(x,h)=y$ admits the expression that 
\begin{equation} \label{sol_for_x}
x=g(y,h):=-v_y(y,h).
\end{equation} 
Defining $f(\cdot,h)$ as the inverse of $g(\cdot,h)$, we have that
\begin{equation}\label{dualrelationship}
u(x,h) = v\big(f(x,h), h\big) + x f(x,h).
\end{equation}
Note that $v$ has different expressions in regions $c =0$, $0 < c < h$ and $c = h$, the function $f$ should also have the piecewise form across these regions. By the definition of $g$ in \eqref{sol_for_x}, the invertibility of the map $x \mapsto g(x,h)$ is guaranteed by the following important result and its proof is deferred to Section \ref{otherproofs}.
\begin{lem}\label{vyy_positive}
In all three regions, we have that $v_{yy}(y, h) > 0$, $\forall h >0$ and the inverse Legendre transform $u(x,h)=\inf_{y\geq(1-\lambda)e^{-(1-\lambda)\beta h}} (v(y,h)+xy)$ is well defined.  Moreover, this implies that the feedback optimal portfolio $\pi^{\dagger}(y,h)>0$ always holds.
\end{lem}

Using \eqref{sol_for_x} and Proposition \ref{dual_value_function}, the function $f$ is implicitly determined in different regions by the following equations:
\begin{itemize}
\item[(i)] If $f(x,h) \geq e^{\lambda \beta h}$, $f(x,h)=f_1(x,h)$ can be determined by
\begin{align*}
x=-C_2(h) r_2  \big(f_1(x,h)\big)^{r_2-1}.
\end{align*}
\item[(ii)] If $e^{(\lambda-1)\beta h} < f(x,h) < e^{\lambda \beta h}$, Lemma \ref{vyy_positive} implies that $v_y(y,h)$ is strictly increasing in $y$ and $f(x,h)=f_2(x,h)$ is uniquely determined by
\begin{align}\label{f2-equt}
x=&-C_3(h) r_1 \big(f_2(x,h)\big)^{r_1-1} - C_4(h) r_2 \big(f_2(x,h)\big)^{r_2-1}\notag \\&
 - \frac{1}{r\beta } \left(\ln f_2(x,h)-\lambda\beta h +\frac{\kappa^2}{2r}\right).
\end{align}
\item[(iii)] If $(1-\lambda)e^{(\lambda-1)\beta h} \leq f(x,h) \leq e^{(\lambda-1)\beta h}$, Lemma \ref{vyy_positive} implies that $v_y(y,h)$ is strictly increasing in $y$ and $f(x,h)=f_3(x,h)$ is uniquely determined by
\begin{align}\label{f3-equt}
x= -C_5(h) r_1 \big(f_3(x,h)\big)^{r_1-1} - C_6(h) r_2 \big(f_3(x,h)\big)^{r_2-1} + \frac{h}{r}.
\end{align}.
\end{itemize}

In region $\mathcal{R}_1$, we can obtain that $f_1(x,h)=( \frac{-x}{C_2(h) r_2} )^{\frac{1}{r_2-1}}$. In addition, $f_1(x,h) \geq e^{\lambda \beta h}$ if and only if $x \leq x_{\text{zero}}(h)$, where we define
\begin{align}\label{x1-def}
x_{\text{zero}}(h):=-e^{\lambda \beta h (r_2-1)} C_2(h)r_2,
\end{align}
which corresponds to the threshold that the optimal consumption becomes zero whenever $x<x_{\text{zero}}(h)$.

In region $\mathcal{R}_2$, the function $f_2$ is uniquely (implicitly) determined by equation \eqref{f2-equt} when $x_{\text{zero}}(h) < x < x_{\text{aggr}}(h)$, where $x_{\text{aggr}}(h)$ is the solution of
$$
f_2(x,h) = e^{(\lambda-1)\beta h}.
$$
In view of \eqref{f2-equt}, we can obtain the boundary explicitly by
\begin{align}\label{x2-def}
x_{\text{aggr}}(h)=-C_3(h) r_1 e^{(\lambda-1)(r_1-1)\beta h}  - C_4(h) r_2 e^{(\lambda-1)(r_2-1)\beta h} + \frac{h}{r}-\frac{\kappa^2}{2r^2\beta},
\end{align}
which corresponds to the threshold that the consumption stays below the historical maximum level whenever $x<x_{\text{aggr}}(h)$.

\begin{rem}\label{discussthresh}
In addition, as in Remark \ref{comparec-lambdah}, we know that the optimal consumption falls below the reference level if and only if $1<f_2(x,h)<e^{\lambda\beta h}$. Using \eqref{f2-equt} again, we can determine the critical point $x_{\text{modr}}(h)$ by
\begin{align*}
x_{\text{modr}}(h):=-C_3(h) r_1 - C_4(h) r_2 + \lambda\frac{h}{r}-\frac{\kappa^2}{2r^2\beta}.
\end{align*}
It then follows that if and only if the wealth level $X_t^*$ satisfies $x_{\text{zero}}(H^*_t)<X^*_t< x_{\text{modr}}(H^*_t)$, the optimal consumption rate meets the moderate plan that $0<c_t^*<\lambda H_t^*$.

\end{rem}
\vspace{3mm}

In region $\mathcal{D}_1\cup\mathcal{D}_2$, the expression of $f_3$ is uniquely determined by the equation \eqref{f3-equt} when $x_{\text{aggr}}(h) \leq x \leq x_{\text{lavs}}(h)$, where $x_3$ is the solution of
\begin{align*}
f_3(x,h) = (1-\lambda) e^{(\lambda-1) \beta h}.
\end{align*}
It follows from \eqref{f3-equt} that the boundary $x_{\text{lavs}}(h)$ is explicitly given by
\begin{align}\label{x3-def}
x_{\text{lavs}}(h):=&-C_5(h)r_1(1-\lambda)^{r_1-1}e^{(\lambda-1)(r_1-1)\beta h} \nonumber \\
&-C_6(h)r_2(1-\lambda)^{r_2-1}e^{(\lambda-1)(r_2-1)\beta h}+\frac{h}{r},
\end{align}
which corresponds to the threshold that the optimal consumption is extremely lavish that $c_t^*$ creates the new maximum level whenever $x=x_{\text{lavs}}(h)$.

Moreover, in view of definitions of $C_5(h)$ and $C_6(h)$ in \eqref{C5h} and \eqref{C6h}, one can check that $x_{\text{lavs}}(h)$ is strictly increasing in $h$ and hence we can define the inverse function
\begin{align}\label{inverseh}
\tilde{h}(x):=(x_3)^{-1}(x),\ \ x\geq 0.
\end{align}
Along the boundary $x=x_{\text{lavs}}(h)$, the feedback form of the optimal consumption in \eqref{feedbackcp} for $y=(1-\lambda) e^{(\lambda-1) \beta h}$ is given by $c^*(x)=\frac{1}{(\lambda-1)\beta} \ln (\frac{1}{1-\lambda} f_3(x,\tilde{h}(x)))$, which only depends on the variable $x$. That is, the optimal consumption can be determined by the current wealth process $X_t^*$ and the associated running maximum process $H_t^*$ is instantly increasing.

In Figure 1 below, we graph all boundary curves $x_{\text{zero}}(h)$, $x_{\text{modr}}(h)$, $x_{\text{aggr}}(h)$ and $x_{\text{lavs}}(h)$ as functions of $h\geq 0$ on the left panel and plot them in terms of the parameter $\lambda\in[0.01,0.98]$ on the right panel (recall that each $C_i(h;\lambda)$ depends on $\lambda$). Although $x_{\text{zero}}(h)$, $x_{\text{modr}}(h)$, $x_{\text{aggr}}(h)$ and $x_{\text{lavs}}(h)$ are complicated nonlinear functions of $h$, the left panel illustrates that all boundary curves are increasing in $h$. This is consistent with the intuition that if the past reference level is higher, the investor would expect larger wealth thresholds to trigger the change of consumption patterns. Recall that we only consider the effective domain that is the region below (and including) the boundary curve $x_{\text{lavs}}(h)$. It is interesting to see from the right panel that $x_{\text{zero}}(1;\lambda)$ and $x_{\text{aggr}}(1;\lambda)$ are decreasing in $\lambda$, while $x_{\text{modr}}(1;\lambda)$ and $x_{\text{lavs}}(1;\lambda)$ are instead increasing in $\lambda$. That is, if the investor clings to a larger proportion of the past spending maximum, it is more likely that the investor will switch from zero consumption to positive consumption (for a low wealth level) and switch from a consumption $c^*<H_t^*$ to the past maximum level $H^*_t$ (for a high wealth level). On the other hand, with a higher proportion $\lambda$, the investor foresees that any aggressive consumption may lead to a ratcheting high reference that will depress all future utilities. As a consequence, the investor will accumulate a larger wealth to change from the moderate consumption $c_t^*<\lambda H_t^*$ to the pattern $c_t^*\geq \lambda H_t^*$ or consume in a way creating a new maximum record that $c_t^*>H_s^*$ for $s<t$, which is consistent with the right panel.

\begin{figure}[h]
\includegraphics[height=1.7in]{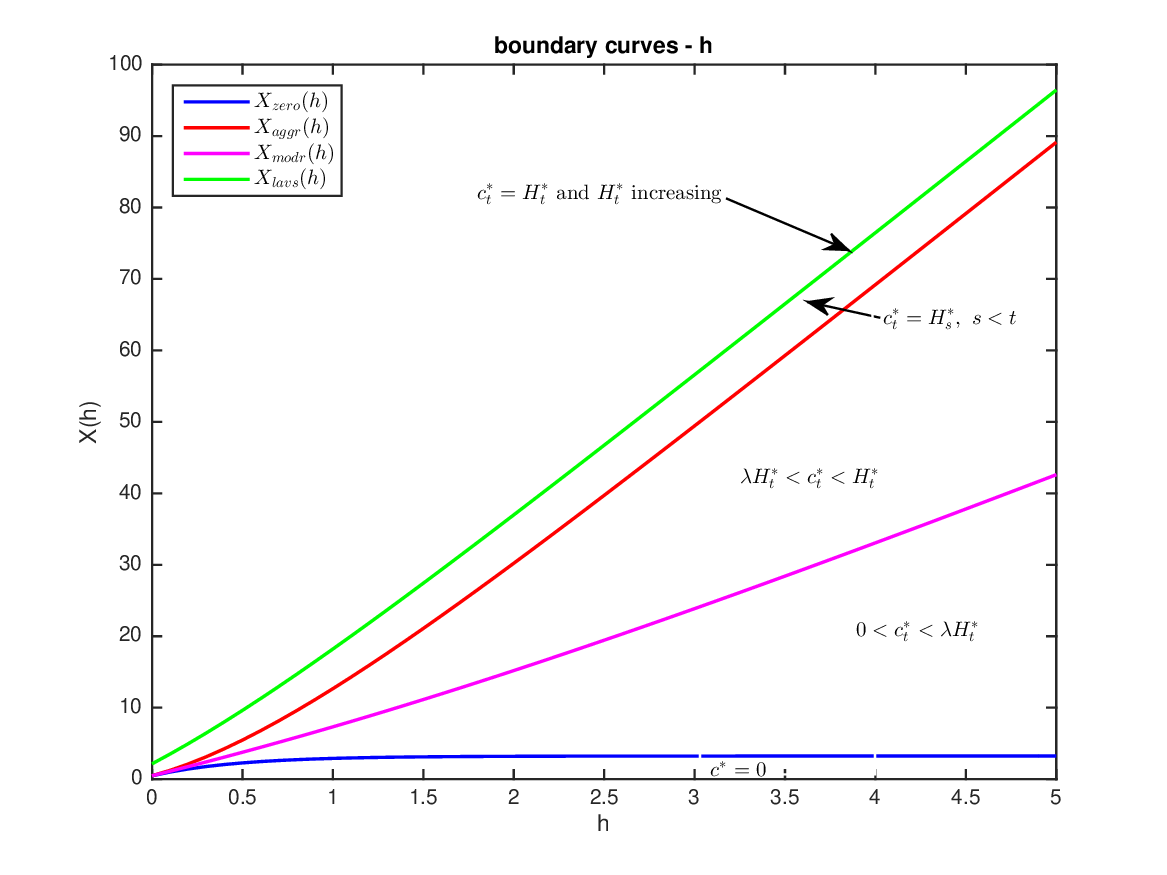}\hspace{0in}\includegraphics[height=1.7in]{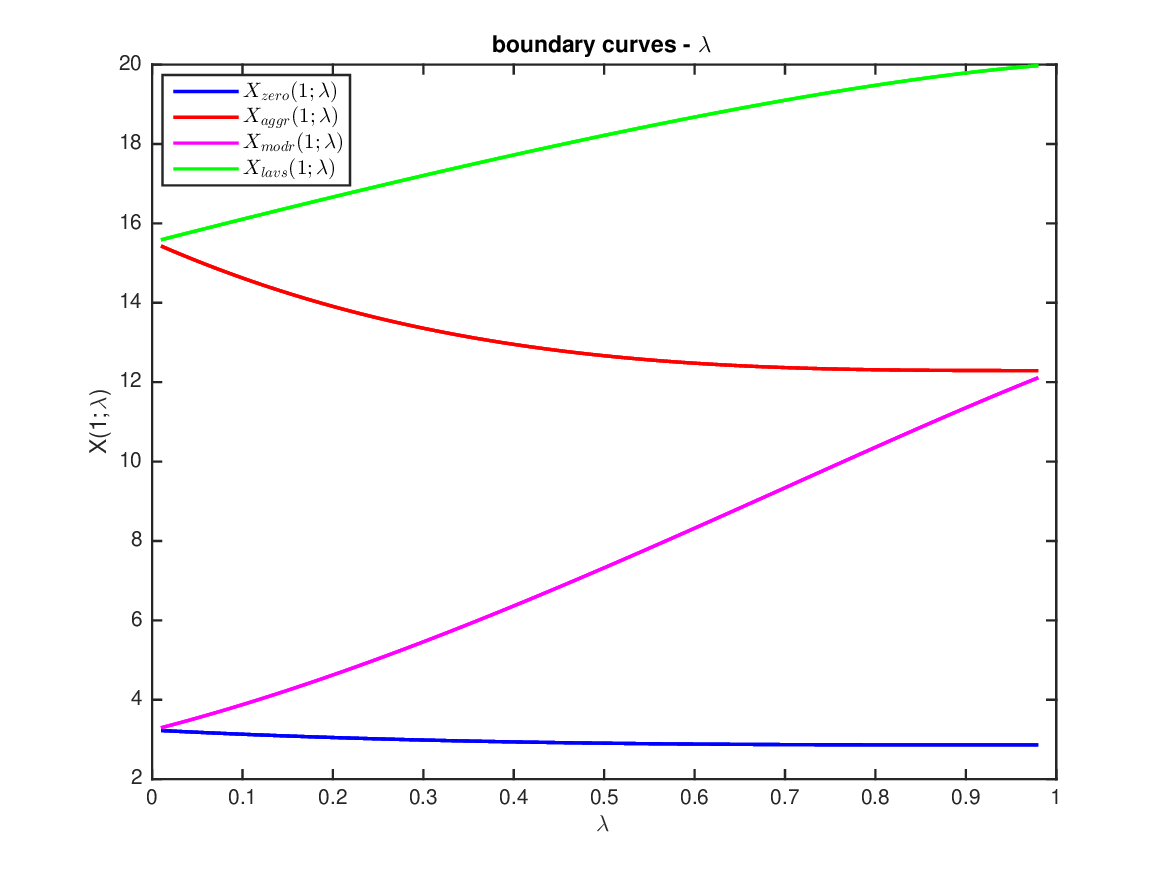}
\caption{\footnotesize{Fix parameters $r = 0.05$, $\mu = 0.1$, $\sigma = 0.25$, $\beta=1$ in both panels. Left panel: graphs of boundary curves in terms of $h$ with $\lambda=0.5$; Right panel: graphs of boundary curves in terms of $\lambda$ with $h=1$.}}
\end{figure}

In particular, the boundary curve $x_{\text{modr}}(1;\lambda)$ in the right panel illustrates that the more the investor cares about the past consumption peak $H_t^*$, the more conservative the investor will become. This may partially explain the real life situations that the constantly aggressive consumption behavior may not result in a long term happiness. A high consumption plan also creates a high level of psychological competition and hence the aggressive consumption behavior may not be sustainable for the life time. A wise investor who takes into account the past reference will strategically lower the consumption rate from time to time (triggered by a wealth threshold) below the dynamic reference such that the reference process can be maintained at a reasonable level and the overall performance can eventually become a win.

Plugging different pieces of $f$ back into equation \eqref{dualrelationship}, we can readily get the next result, in which the value function $u$ and optimal feedback controls are all given in terms of the primal variables $x$ and $h$, and the existence of the unique strong solution to SDE \eqref{originw} under optimal controls can be obtained.

\begin{cor} \label{value_function_primal}
For $(x,h)\in\mathcal{C}$ and $0<\lambda<1$, let us define the piecewise function
\begin{align*}
& f(x,h)=\left\{
\begin{aligned}
& \big( -x/C_2(h) r_2 \big)^{\frac{1}{r_2-1}},   & & \mbox{if } x \leq x_{\text{zero}}(h), \\
& f_2(x,h),  & & \mbox{if } x_{\text{zero}}(h) < x < x_{\text{aggr}}(h), \\
& f_3(x,h),  & & \mbox{if } x_{\text{aggr}}(h) \leq x \leq x_{\text{lavs}}(h),\\
\end{aligned}
\right.
\end{align*}
where $f_2(x,h)$ and $f_3(x,h)$ are determined by \eqref{f2-equt} and \eqref{f3-equt}. The value function $u(x,h)$ of the control problem in \eqref{primalvalue} is given by
\begin{align*}
& u(x,h) \nonumber \\
=&\left\{
\begin{aligned}
&  C_2(h) f(x,h)^{r_2} - \frac{1}{r\beta} e^{\lambda \beta h} + x f(x,h),  & & \mbox{if } x \leq x_{\text{zero}}(h), \\
& C_3(h) \big(f(x,h)\big)^{r_1} + C_4(h) \big(f(x,h)\big)^{r_2} & & \\
& +\frac{f(x,h)}{r\beta } \left(\ln f(x,h)-\lambda\beta h+\frac{\kappa^2}{2r} -1+xr\beta \right),  & & \mbox{if } x_{\text{zero}}(h) < x < x_{\text{aggr}}(h), \\
& C_5(h) \big(f(x,h)\big)^{r_1} + C_6(h) \big(f(x,h)\big)^{r_2} -\frac{1}{r}h f(x,h) & &\\
&-\frac{1}{r\beta} e^{(\lambda-1)\beta h} + x f(x,h),  & & \mbox{if } x_{\text{aggr}}(h) \leq x \leq x_{\text{lavs}}(h).\\
\end{aligned}
\right.
\end{align*}
To distinguish the feedback functions $c^{\dagger}(y,h)$ and $\pi^{\dagger}(y,h)$ in \eqref{feedbackcp} and \eqref{feedbackpi} based on dual variables, let us denote $c^{*}(x,h)$ and $\pi^{*}(x,h)$ as the feedback functions of the optimal consumption and portfolio using primal variables $(x,h)$. We have that $c_t^*=c^{*}(X_t^*, H_t^*)$ and $\pi_t^*=\pi^{*}(X_t^*, H_t^*)$ where 
\begin{align}\label{consumption:primal}
c^{*}(x,h) =\left\{
\begin{aligned}
&0, & & \mbox{if } x \leq x_{\text{zero}}(h), \\
&- \frac{1}{\beta}\ln f(x,h) + \lambda h,  & & \mbox{if } x_{\text{zero}}(h) < x < x_{\text{aggr}}(h), \\
&h,  & & \mbox{if } x_{\text{aggr}}(h) \leq x < x_{\text{lavs}}(h), \\
&\frac{1}{(\lambda-1)\beta} \ln \Big(\frac{1}{1-\lambda} f\big(x,\tilde{h}(x)\big)\Big),     & & \mbox{if } x= x_{\text{lavs}}(h),
\end{aligned}
\right.
\end{align}
where $\tilde{h}(x)$ is given in \eqref{inverseh}, and
\begin{align}\label{invest:primal}
& \pi^{*}(x,h) =\frac{\mu-r}{\sigma^2}\nonumber \\
&\times\left\{
\begin{aligned}
&(1-r_2)x, & & \mbox{if } x \leq x_{\text{zero}}(h), \\
&\frac{2r}{\kappa^2}C_3(h)f^{r_1-1}(x,h)+\frac{2r}{\kappa^2}C_4(h)f^{r_2-1}(x,h)+\frac{1}{r\beta},  & & \mbox{if } x_{\text{zero}}(h) < x < x_{\text{aggr}}(h), \\
& \frac{2r}{\kappa^2}C_5(h) f^{r_1-1}(x,h) + \frac{2r}{\kappa^2}C_6(h) f^{r_2-1}(x,h),  & & \mbox{if } x_{\text{aggr}}(h) \leq x \leq x_{\text{lavs}}(h).
\end{aligned}
\right.
\end{align}
We have that $0<c_t^{*}(X_t^*,H_t^*)<\lambda H_t^{*}$ if and only if $x_{\text{zero}}(H_t^*)< X_t^*< x_{\text{modr}}(H_t^*)$. 

Moreover,  for any initial value $(X_0^*, H_0^*) = (x, h) \in \mathcal{C}$, the stochastic differential equation
\begin{align}\label{wealthSDE}
d X_t^* =  r X_t^* dt+ \pi^*(\mu -r)dt+ \pi^* \sigma dW_t- c^*dt
\end{align}
has a unique strong solution given the optimal feedback control $(c^*, \pi^*)$ as above.
\end{cor}

Based on Corollary \ref{value_function_primal}, we can readily obtain the next result of the asymptotic behavior of the optimal consumption-wealth ratio $c_t^*/X_t^*$ and the investment amount $\pi^*_t$ when the wealth is sufficiently large. The proof is given in Section \ref{otherproofs}.
\begin{cor}\label{propasymp}
For $0<\lambda<1$, as $x\leq x_{\text{lavs}}(h)$, the asymptotic behavior of large wealth $x\to+\infty$ is equivalent to $\lim_{h\to+\infty} x_{\text{lavs}}(h)=+\infty$ thanks to the explicit expression of $x_{\text{lavs}}(h)$ in \eqref{x3-def}. We then have that
\begin{align*}
\lim_{h\to+\infty} \frac{c^*(x_{\text{lavs}}(h), h)}{ x_{\text{lavs}}(h)}=r, \ \ \ \lim_{h\to+\infty} \pi^*(x_{\text{lavs}}(h), h) = \frac{(\mu-r)(1-\lambda)^{r_1-1}}{r\beta\sigma^2}.
\end{align*}
As the wealth level gets sufficiently large, the optimal consumption is asymptotically proportional to the wealth level that $c_t^{*}\approx rX_t^{*}$ and the optimal investment converges to a constant level that $\pi^{*}_t\approx (\mu-r)\frac{(1-\lambda)^{r_1-1}}{r\beta\sigma^2}$. That is, the investor will only allocate a constant amount of wealth into the risky asset and save most of his wealth in the bank account.
\end{cor}

\subsection{Comparison with Some Related Works}
Given the feedback optimal controls in \eqref{consumption:primal} and \eqref{invest:primal}, we briefly present here some comparison results with Arun \cite{Arun} and Guasoni et al. \cite{GHR} on the optimal consumption affected by the past spending maximum. We stress that Arun \cite{Arun} considers the optimal consumption under a standard time separable power utility that $\sup_{(c,\pi)}\mathbb{E}[\int_0^{\infty}e^{-\rho t}\frac{(c_t)^p}{p}dt]$, while the drawdown constraint that $c_t\geq \lambda H_t^*$, $0<\lambda<1$, is only imposed in the set of admissible controls. See also Dybvig \cite{Dyb} with the ratcheting constraint for $\lambda=1$. A similar optimal dividend control problem with the drawdown constraint is also formulated and studied in Angoshtari et al. \cite{BAY}. On the other hand, Guasoni et al. \cite{GHR} studies the optimal consumption under a Cobb-Douglas utility that $\sup_{(c,\pi)}\mathbb{E} [\int_0^{\infty}\frac{(c_t/H_t^{\alpha})^p}{p}dt ]$, where the utility is defined on the ratio of the consumption rate and the consumption running maximum. By virtue of the power utility on consumption rate, the optimal consumption in Arun \cite{Arun} and Guasoni et al. \cite{GHR} automatically satisfy $c_t^*\geq 0$. On the other hand, we are interested in the exponential utility that measures the difference between the consumption rate and the consumption running maximum process. The non-negativity constraint $c_t\geq 0$ needs to be taken care of in solving the HJB equation. As our utility differs from Arun \cite{Arun} and Guasoni et al. \cite{GHR}, the feedback optimal consumption is certainly distinct from their results. But we can compare the optimal consumption behavior when the wealth level becomes extremely low and extremely high. These major differences are summarized in Table 1 as below.

\begin{table}[ht]
\centering
\begin{tabular}{|c | c| c|}
\hline
 & when the wealth is low & when the wealth is high \\
\hline
Arun \cite{Arun} & \begin{tabular}{@{}c@{}} the problem is only well defined \\
if the wealth level satisfies\\ the constraint $X_t^*\geq \frac{a}{r}H_t^*$;\\ moreover, $c_t^*=aH_t^*$ for some $a<1$,\\ i.e., the optimal consumption is\\ proportional to running maximum \end{tabular} & \begin{tabular}{@{}c@{}} $c_t^*=bX_t^*$ for some $b>0$,\\ i.e., the optimal consumption is\\ proportional to optimal wealth \end{tabular} \\
\hline
Guasoni et al. \cite{GHR}  &  \begin{tabular}{@{}c@{}} when $X^*_t\leq aH_t^*$,\\ $c_t^*=\frac{1}{a}X_t^*$ for some $a>0$,\\ i.e., the optimal consumption is\\ proportional to optimal wealth  \end{tabular} & \begin{tabular}{@{}c@{}} when $X^*_t\geq bH_t^*$,\\ $c_t^*=\frac{1}{b}X_t^*$ for some $b>0$,\\ i.e., the optimal consumption is\\ proportional to optimal wealth   \end{tabular} \\
\hline
The present paper & \begin{tabular}{@{}c@{}} when $X^*_t\leq x_{\text{zero}}(H_t^*)$,\\ $c_t^*=0$ even when $X_t^*>0$\\  \end{tabular}  & \begin{tabular}{@{}c@{}} when $X_t^*=x_{\text{lavs}}(H_t^*)$,\\ $c_t^*$ is a nonlinear function of $X_t^*$,\\ as $X^*_t\rightarrow+\infty$,\\ $c_t^*\approx r X_t^*$ that is asymptotically\\ proportional to the optimal wealth  \end{tabular} \\
\hline
\end{tabular}
\caption{Comparison results on optimal consumption when the wealth level is extremely small and extremely large}
\end{table}

In addition, in Arun \cite{Arun} and Guasoni et al. \cite{GHR} (see also Angoshtari et al. \cite{BAY}), one can change variables and focus on the new state process $X_t/H_t$ to reduce the dimension. As a consequence, all thresholds for the wealth variable $x$ separating different regions for the piecewise optimal control in these works are simply linear functions of $h$. In contrary, our utility is defined on the difference $c_t-\lambda H_t$, and the change of variables is no longer applicable. The HJB equation is genuinely two dimensional that complicates the characterization of all boundary curves separating different regions. We choose to apply the dual transform to $x$ and treat $h$ as a parameter in the whole analysis. The smooth fit principle and inverse transform can help us to identify these boundary curves fully explicitly. We finally can express these thresholds $x_{\text{zero}}(h)$, $x_{\text{aggr}}(h)$ and $x_{\text{lavs}}(h)$ in \eqref{x1-def}, \eqref{x2-def} and \eqref{x3-def} as nonlinear functions of $h$ (see the left panel of Figure 1), which are much more complicated than their counterparts in Arun \cite{Arun} and Guasoni et al. \cite{GHR}.

\section{Numerical Examples and Sensitivity Analysis}\label{sec:numerics}
We present here some numerical examples of sensitivity analysis on model parameters using the closed-form value function and feedback optimal controls in Corollary \ref{value_function_primal} and discuss some interesting financial implications.

Let us first examine the sensitivity with respect to the weight parameter $0<\lambda<1$ in Figure 2 by plotting some comparison graphs of the value function, the feedback optimal consumption and the feedback optimal portfolio. From the middle panel, we can see again that $x_{\text{zero}}(1;\lambda)$ and $x_{\text{aggr}}(1;\lambda)$ are decreasing in $\lambda$ but $x_{\text{lavs}}(1;\lambda)$ is increasing in $\lambda$. More importantly, for each fixed $x$ such that $x_{\text{zero}}(1; \lambda_{\text{max}})<x<x_{\text{aggr}}(1;\lambda_{\text{min}})$, the feedback optimal consumption $c^*(x,1;\lambda)$ is increasing in the parameter $\lambda\in (\lambda_{\text{min}}, \lambda_{\text{max}})$, which matches with the intuition that a higher reference weight parameter will induce a higher consumption. However, the middle panel also illustrates that this intuition is only partially correct as it only holds when the consumption does not surpass the historical maximum. When the wealth level gets higher, the investor can freely choose to consume in a lavish way. Then, for $x>x_{\text{lavs}}(1;\lambda_{\text{min}})$, we can see that a smaller $\lambda$ leads to an earlier lavish consumption $c^*(x,1)>h=1$ creating a new $h$. But when the wealth continues to increase, the consumption with a larger $\lambda$ will eventually dominate its counterpart with a smaller $\lambda$.

We can also observe from the right panel of Figure 2 that for the fixed $x>0$, $\pi^*(x,1;\lambda)$ is decreasing in $\lambda$, which is consistent with the middle panel that the optimal consumption level is lifted up by a larger value of $\lambda$. When the capital is sufficient, the investor may strategically invest less in the market to save more cash to support the higher consumption plan induced by the larger $\lambda$. The left panel of Figure 2 further shows that the value function $u(x,h;\lambda)$ is actually decreasing in $\lambda$. Note that our utility is measured by the difference of the consumption $c_t^*$ and the reference process $\lambda H_t^*$. When $\lambda$ increases, both $c^*_t$ and $\lambda H_t^*$ increase. We can see from the left panel that $\lambda H^*_t$ increases faster than the consumption $c^*_t$ during the life cycle that leads to a drop of $c^*_t-\lambda H_t^*$ and the decline in the value function. It is also interesting to observe that when $\lambda$ is large, for $x_{\text{aggr}}(1;\lambda)<x<x_{\text{lavs}}(1;\lambda)$, the optimal portfolio $\pi^*(x,1;\lambda)$ may decrease even when the wealth $x$ increases. But when $x$ continues to increase such that $x>x_{\text{lavs}}(1;\lambda)$, we can see that the optimal portfolio $\pi^*(x,1;\lambda)$ starts to increase in $x$. That the optimal portfolio might be decreasing in $x$ differs from some existing works and it is a consequence of our specific path-dependent preference. Indeed, when the wealth is sufficient to support the aggressive consumption that $c_t^*>\lambda H_t^*$ but the reference process $\lambda H_t^*$ is not changed, the investor may strategically withdraw the portfolio amount from the financial market to support the consumption plan. This non-standard phenomenon is more likely to happen when the reference parameter $\lambda$ is large such that the resulting consumption $c_t^*>\lambda H_t^*$ is very high (see Figure 2), or when the risky asset performance is not good enough (low return as in Figure 3 or high volatility as in Figure 4). However, when $x$ gets abundant such that the investor starts to increase the reference level $\lambda H_t^*$, the large amount of consumption will quickly eat the capital, and the investor can no longer compromise the portfolio amount to support consumption. It turns to be optimal for the investor to increase the portfolio and accumulate more wealth from the financial market to sustain the extremely high consumption decision. We stress that the non-standard phenomenon that $\pi^*(x,1)$ may decrease in $x$ is consequent on some complicated trade-offs of all model parameters. Under some appropriate model parameters, the optimal portfolio $\pi^*(x,1)$ is always increasing in $x>0$, which matches with the intuition that we will invest more if we have more.

\begin{figure}[h]
\hspace{-0.6in}
\includegraphics[height=1.35in]{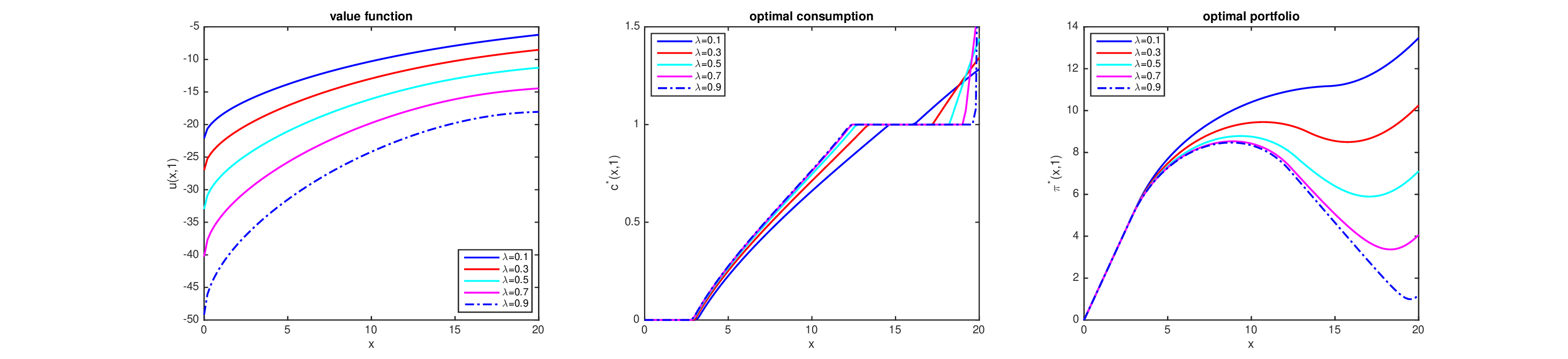}
\vspace{-0.2in}
\caption{\footnotesize{Fix parameters $r =0.05$, $\mu = 0.1$, $\sigma = 0.25$, $\beta=1$ and the variable $h=1$. With changes of $\lambda=0,1, 0.2, 0.5, 0.7$ and $0.9$, we plot graphs of the value function $u(x,1)$ (left panel), the optimal consumption $c^*(x,1)$ (middle panel) and the optimal portfolio $\pi^*(x,1)$ (right panel) for $x\in[0,20]$.}}
\end{figure}
\vspace{-0.1in}

We next discuss the impact of the drift parameter $\mu$ based on plots in Figure 3. Firstly, we can see from the left panel that both $x_{\text{zero}}(1;\mu)$ and $x_{\text{aggr}}(1;\mu)$ are decreasing in $\mu$. That is, the higher return the risky asset has, the less wealth that the investor needs to start the positive consumption and initiate the lavish consumption to increase the reference process. Moreover, the feedback optimal consumption is also increasing in $\mu$. These observations are consistent with the real life situation that the bull market will help the investor to accumulate more wealth so that the investor will become more optimistic to develop a more aggressive consumption pattern. Secondly, as one can expect, the right panel of Figure 3 illustrates that the optimal portfolio in the financial market increases as the return increases. In addition, the left panel shows that the primal value function is increasing in $\mu$. It illustrates that when the return $\mu$ increases, the increment in optimal consumption rate $c^*_t$ dominates the increment in the reference process $\lambda H^*_t$ so that the value function is lifted up. Thirdly, combining Figure 2 and Figure 3, for the same wealth level $x$, we know that the optimal portfolio $\pi^*(x,1;\lambda, \mu)$ is decreasing in $\lambda$ but increasing in $\mu$. As a consequence, for those investors who are more addictive to the past reference level, the market premiums need to be sufficiently high to attract them to invest in the risky asset. This observation may partially explain the observed \textit{equity premium puzzle} (see Mehra and Prescott \cite{Meh} and many subsequent works) from the perspective of our proposed path-dependent utility with past spending maximum.

\begin{figure}[h]
\hspace{-0.6in}
\includegraphics[height=1.35in]{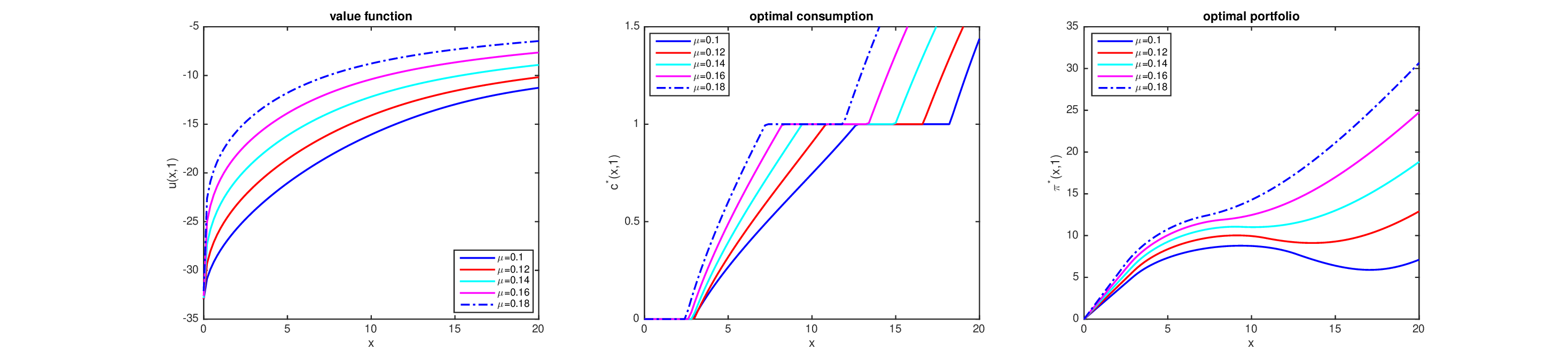}
\vspace{-0.2in}
\caption{\footnotesize{Fix parameters $r = 0.05$, $\sigma = 0.25$, $\lambda = 0.5$, $\beta=1$ and the variable $h = 1$. With changes of $\mu=0.10$, $0.12$, $0.14$, $0.16$ and $0.18$, we plot graphs of the value function $u(x,1)$ (left panel), the optimal consumption $c^*(x,1)$ (middle panel) and the optimal portfolio $\pi^*(x,1)$ (right panel) for $x\in[0,20]$. }}
\end{figure}
\vspace{-0.2in}

At last, we perform sensitivity with respect to the volatility $\sigma$ in Figure $4$. From the middle panel of Figure 4, we observe that the monotonicity of thresholds $x_{\text{zero}}(1;\sigma)$ and the optimal consumption $c^*(x,1;\sigma)$ on the parameter $\sigma$ do not hold in general and become much subtle and complicated. Only when the wealth level is sufficiently large, the optimal consumption $c^*(x,1;\sigma)$ is decreasing in $\sigma$. It is only clear that the threshold $x_{\text{aggr}}(1;\sigma)$ is increasing in $\sigma$. This observation can be explained that when the wealth level is sufficiently high, the less volatile the risky asset is, the more optimistically the investor will behave in wealth management and consumption plan. In other words, the investor will consume more when $\sigma$ is smaller and lower the threshold to start some large expenditures such that the spending maximum is increased. However, when the wealth level is too low, the investor will become more conservative towards the risky asset account and rely more on interest rate to accumulate enough wealth to initiate a positive consumption. As a consequence, the threshold $x_{\text{zero}}(1;\sigma)$ is not necessarily monotone in $\sigma$. The left and right panels of Figure 4 also show that both the value function and the optimal portfolio are decreasing in the parameter $\sigma$. These graphs are consistent with the real life observations that if the risky asset has a higher volatility, the investor allocates less wealth in the risky asset and the life cycle value function also becomes lower.

\begin{figure}[h]
\hspace{-0.6in}
\includegraphics[height=1.35in]{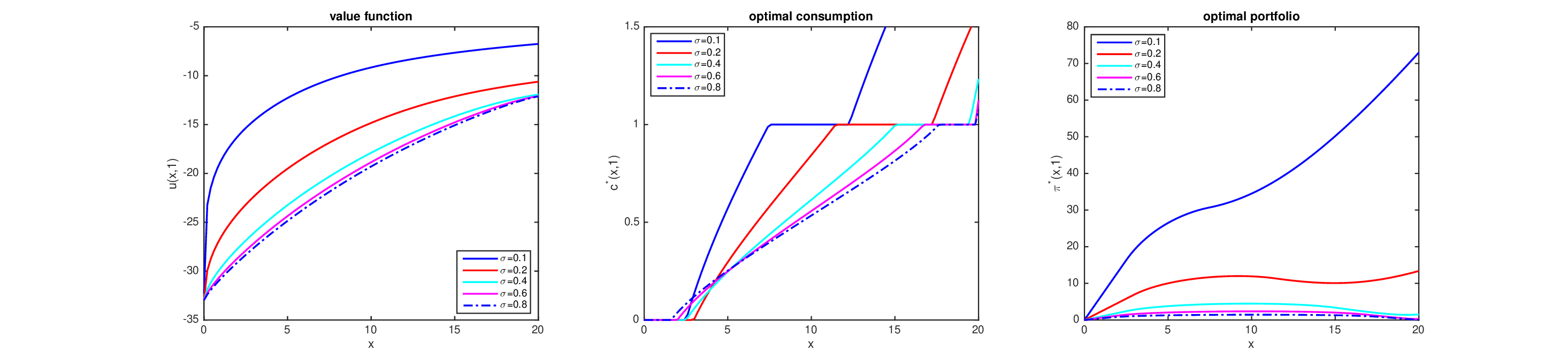}
\vspace{-0.2in}
\caption{\footnotesize{Fix parameters $r = 0.05$, $\mu = 0.1$, $\beta=1$, $\lambda=0.5$ and the variable $h=1$. With changes of $\sigma=0.1, 0.2, 0.4, 0.6, 0.8$, we plot graphs of the value function $u(x,1)$ (left panel), the optimal consumption $c^*(x,1)$ (middle panel) and the optimal portfolio $\pi^*(x,1)$ (right panel) for $x\in[0,20]$. }}
\end{figure}

\section{Proofs of Main Results}\label{sec:proofs}

\subsection{Proofs of some auxiliary results in Section \ref{sec:solveHJB} }\label{otherproofs}

\begin{proof}[Proposition \ref{dual_value_function}]
We can first obtain the special solution $v(y,h)=-\frac{1}{r\beta} e^{\lambda\beta h}$ for the first equation, then $v(y,h)=-\frac{y}{r\beta}+\frac{y}{r\beta} (\ln y-\lambda\beta h+\frac{\kappa^2}{2r} )$ for the second equation, and finally $v(y,h)=-\frac{1}{r}hy-\frac{1}{r\beta} e^{(\lambda-1)\beta h}$ for the third equation in \eqref{EDPtildeU2}. Therefore, we can summarize the general solution of the ODE \eqref{EDPtildeU2} by
\begin{align}\label{explicitdual}
&v(y,h) = \nonumber \\ &\left\{
\begin{aligned}
& C_1(h) y^{r_1} + C_2(h) y^{r_2} - \frac{1}{r\beta} e^{\lambda \beta h}, \mbox{if } y \geq e^{\lambda \beta h}, \\
& C_3(h) y^{r_1} + C_4(h) y^{r_2}-\frac{y}{r\beta}+\frac{y}{r\beta } \left(\ln y-\lambda\beta h+\frac{\kappa^2}{2r} \right), \mbox{if } e^{(\lambda-1)\beta h} < y < e^{\lambda \beta h}, \\
& C_5(h) y^{r_1} + C_6(h) y^{r_2} -\frac{1}{r}hy-\frac{1}{r\beta} e^{(\lambda-1)\beta h}, \mbox{if } (1-\lambda)e^{(\lambda-1)\beta h} \leq y \leq e^{(\lambda-1)\beta h},\\
\end{aligned}
\right.
\end{align}
in which $C_i(h)$, $i=1,...,6,$ are functions of $h$ to be determined.

By the explicit form of $v(y,h)$ in \eqref{explicitdual} along the free boundary $y=(1-\lambda)e^{(\lambda-1)\beta h}$, the condition $v_h(y,h)=0$ in \eqref{freedul} implies that
\begin{equation} \label{eq:boundary}
C_5'(h) (1-\lambda)^{r_1} e^{(\lambda-1)\beta hr_1} + C_6'(h) (1-\lambda)^{r_2} e^{(\lambda-1) \beta hr_2} = 0.
\end{equation}

Similar to the case when $\lambda=0$, the free boundary condition $v_y(y,h)\rightarrow 0$ in \eqref{yhcond-2} implies that $y\rightarrow +\infty$. Together with free boundary conditions in \eqref{yhcond-2} and the formula of $v(y,h)$ in the region $y\geq e^{\lambda\beta h}$, we deduce that $C_1(h)\equiv 0$. Moreover, it is easy to see that as $h\rightarrow+\infty$, we get $y\rightarrow 0$ in the third region $(1-\lambda)e^{(\lambda-1)\beta h} \leq y \leq e^{(\lambda-1)\beta h}$ and therefore the boundary conditions in \eqref{yhcond-1} also implies the asymptotic condition that $C_6(h) \rightarrow 0$ as $h\rightarrow+\infty$.

To determine the remaining parameters, we apply the smooth-fit conditions with respect to the variable $y$ at the two boundary points $y=e^{\lambda\beta h}$ and $y=e^{(\lambda-1)\beta h}$. After simple manipulations, we can deduce the system of equations:
\begin{equation*}
\left\{
\begin{aligned}
&C_2(h) e^{\lambda \beta h r_2} =C_3(h) e^{\lambda\beta h r_1} + C_4(h) e^{\lambda \beta h r_2} +\frac{1}{2r^2\beta}e^{\lambda\beta h} \kappa^2, \\
&C_2(h) r_2 e^{\lambda \beta h r_2} =C_3(h) r_1 e^{\lambda\beta h r_1} + C_4(h) r_2 e^{\lambda\beta h r_2} +\frac{1}{2r^2\beta}e^{\lambda\beta h} \kappa^2 , \\
&C_3(h) e^{(\lambda-1)\beta h r_1} + C_4(h) e^{(\lambda-1)\beta h r_2} +\frac{1}{2r^2\beta}e^{(\lambda-1)\beta h} \kappa^2 \\
&= C_5(h) e^{(\lambda-1) \beta h r_1} + C_6(h) e^{(\lambda-1)\beta h r_2} , \\
& C_3(h) r_1 e^{(\lambda-1) \beta h r_1} + C_4(h) r_2 e^{(\lambda-1) \beta h r_2}   +\frac{1}{2r^2\beta}e^{(\lambda-1)\beta h} \kappa^2 \\
&=C_5(h) r_1 e^{(\lambda-1) \beta h r_1} + C_6(h) r_2 e^{(\lambda-1) \beta h r_2}.
\end{aligned}
\right.
\end{equation*}

The system of equations can be solved explicitly. To this end, the linear system can be regarded as linear equations in terms of variables $C_3(h)$, $C_2(h)-C_4(h)$, $C_4(h)-C_6(h)$ and $C_3(h)-C_5(h)$. We can solve the first two equations and obtain $C_3(h)$ explicitly in \eqref{C3h} and $C_2(h)-C_4(h)$. By solving the last two equations, we also get $C_3(h)-C_5(h)$, which yields $C_5(h)$ in \eqref{C5h} by substituting the function $C_3(h)$.

Plugging the derivative $C_5'(h)$ back into the boundary condition \eqref{eq:boundary}, we obtain that
$$
\begin{aligned}
C_6'(h) (1-\lambda)^{r_2} e^{(\lambda-1)\beta h r_2} =& (1-\lambda)^{r_1} e^{(\lambda-1)\beta h r_1} \frac{(r_2-1) \kappa^2}{2(r_1-r_2)\beta r^2}  \\
&\times \left( (\lambda-1)(1-r_1) e^{(\lambda-1)(1-r_1)\beta h}-\lambda (1-r_1) e^{\lambda (1-r_1)\beta h}\right).
\end{aligned}
$$

By using the asymptotic condition that $C_6(h) \rightarrow 0$ when $h \rightarrow + \infty$ and the condition that $\lambda(1-r_2)-(r_1-r_2) < 0$, we can integrate the equation above on both sides, and get $C_6(h)$ explicitly in \eqref{C6h}.

Substituting $C_6(h)$ back to
$$
(r_1-r_2)\big(C_6(h)-C_4(h)\big) e^{(\lambda-1)\beta hr_2}=   (r_1-1) \frac{e^{(\lambda-1)\beta h}}{2r^2\beta} \kappa^2,
$$
we can get $C_4(h)$ in \eqref{C4h}. Substituting $C_4(h)$ to the equation that
$$
(r_1-r_2)\big(C_2(h)-C_4(h)\big) e^{\lambda \beta h r_2} = (r_1-1) \frac{e^{\lambda \beta h}}{2r^2\beta} \kappa^2,
$$
we can at last obtain $C_2(h)$ in \eqref{C2h}.
\end{proof}

\ \\
\begin{proof}[Lemma \ref{vyy_positive}]
We shall analyze each region separately.
\vspace{3mm}

(i) In the region $y \geq e^{\lambda \beta h}$, we have $v_{yy}(y,h) = r_2(r_2-1)C_2(h) y^{r_2-2}$ as $r_2(r_2-1)=\frac{2r}{\kappa^2}>0$ and $C_2(h)>0$ thanks to its expression \eqref{C2h}. The conclusion holds trivially.

\vspace{3mm}

(ii) In the region $(1-\lambda)e^{(\lambda-1)\beta h} \leq y \leq e^{(\lambda-1)\beta h}$, we recall that 
\begin{align*}
v_{yy}(y,h) = r_1(r_1-1)C_5(h) y^{r_1-2} + r_2(r_2-1)C_6(h) y^{r_2-2}.
\end{align*} 
The conclusion easily follows from the fact that $C_5(h)>0$, $C_6(h)>0$ and the identity that $r_1(r_1-1)=r_2(r_2-1)= \frac{2r}{\kappa^2}>0$.

\vspace{3mm}

(iii) In the region $e^{(\lambda-1)\beta h} < y < e^{\lambda \beta h}$, we proceed by the following two steps:

\vspace{3mm}

\textit{Step 1}: To show $v_{yy}(y,h)>0$, it is equivalent to check that $yv_{yy}(y,h)>0$. Let us first show this at the two endpoints $e^{(\lambda-1)\beta h}$ and $e^{\lambda\beta h}$. So at $y=e^{(\lambda-1)\beta h}$ and $y=e^{\lambda\beta h}$, we need to prove the inequality
\begin{equation} \label{eq:inter_inequality1}
\frac{2r}{\kappa^2} y^{r_2-1} \big( C_3(h) y^{r_1-r_2} + C_4(h) \big) + \frac{1}{\beta r} > 0.
\end{equation}
By $C_3(h)$ and $C_4(h)$ in the explicit form, at the point $e^{\lambda \beta h}$, \eqref{eq:inter_inequality1} boils down to proving that
\begin{align*}
& e^{\lambda \beta h (r_2-1)} \frac{1}{(r_1-r_2) \beta} \Bigg( e^{\lambda \beta h (1-r_2)} \frac{r_2-1}{r}\\& + e^{(\lambda-1) \beta h (1-r_2)} \frac{1-r_1}{r} + \frac{(r_2-1) (1-\lambda)^{r_1-r_2}}{r}  \\
& \times \left( \frac{1-r_1}{1-r_2} e^{(\lambda-1)(1-r_2)\beta h} -\frac{\lambda(1-r_1)}{\lambda(1-r_2)-(r_1-r_2)} e^{\left( \lambda(1-r_2)-(r_1-r_2) \right) \beta h} \right) \Bigg)+\frac{1}{r \beta} > 0.
\end{align*}
Using the fact that $e^{\lambda \beta h (1-r_2)} > e^{(\lambda-1) \beta h (1-r_2)}$, we can see that the above is larger than
\begin{align*}
& e^{\lambda \beta h (r_2-1)} \frac{1}{(r_1-r_2) \beta} \Big( e^{\lambda \beta h (1-r_2)} \frac{r_2-1}{r}+ e^{\lambda \beta h (1-r_2)} \frac{1-r_1}{r} \Big) +\frac{1}{r \beta} - e^{\lambda \beta h (r_2-1)}  \\
&\times \frac{(1-r_2) (1-\lambda)^{r_1-r_2}}{(r_1-r_2) \beta r} \\&\times\left( \frac{1-r_1}{1-r_2} e^{(\lambda-1)(1-r_2)\beta h} -\frac{\lambda(1-r_1)}{\lambda(1-r_2)-(r_1-r_2)} e^{\left( \lambda(1-r_2)-(r_1-r_2) \right)\beta h} \right)\\
&=- e^{\lambda \beta h (r_2-1)} \frac{(1-r_2) (1-\lambda)^{r_1-r_2}}{(r_1-r_2) \beta r}\\&\quad\times \Bigg( \frac{1-r_1}{1-r_2} e^{(\lambda-1)(1-r_2)\beta h} -\frac{\lambda(1-r_1)}{\lambda(1-r_2)-(r_1-r_2)} e^{\left( \lambda(1-r_2)-(r_1-r_2) \right)\beta h} \Bigg),
\end{align*}
which is strictly positive. Hence, we have that $y v_{yy}(y,h)>0$ at $y=e^{\lambda \beta h}$.

To show \eqref{eq:inter_inequality1} at the endpoint $y=e^{(\lambda-1)\beta h}$, it is enough to show that
\begin{align*}
&e^{(\lambda-1) \beta h (r_2-1)} \frac{1}{(r_1-r_2) \beta} \Bigg( e^{(\lambda(1-r_2)-(r_1-r_2)) \beta h } \frac{r_2-1}{r}\\&  + e^{(\lambda-1) \beta h (1-r_2)} \frac{1-r_1}{r} + \frac{(r_2-1) (1-\lambda)^{r_1-r_2}}{r} \\
& \times \left( \frac{1-r_1}{1-r_2} e^{(\lambda-1)(1-r_2)\beta h} -\frac{\lambda(1-r_1)}{\lambda(1-r_2)-(r_1-r_2)} e^{\left( \lambda(1-r_2)-(r_1-r_2) \right)\beta h} \right) \Bigg)+\frac{1}{r \beta} > 0.
\end{align*}
By the fact $e^{(\lambda(1-r_2)-(r_1-r_2)) \beta h } < e^{(\lambda-1) \beta h (1-r_2)}$ and similar calculations for $y=e^{\lambda \beta h}$, we can also show that the above term is strictly larger than
\begin{align*}
&- e^{ (\lambda-1) \beta h (r_2-1)}  \frac{(1-r_2) (1-\lambda)^{r_1-r_2}}{(r_1-r_2) \beta r} \\& \times\left( \frac{1-r_1}{1-r_2} e^{(\lambda-1)(1-r_2)\beta h} -\frac{\lambda(1-r_1)}{\lambda(1-r_2)-(r_1-r_2)} e^{\left( \lambda(1-r_2)-(r_1-r_2) \right)\beta h} \right)>0,
\end{align*}
and hence is strictly positive.

\vspace{3mm}

\textit{Step 2}:  In this step, we show that the function
$$
\gamma(y):=y v_{yy}(y,h)= \frac{2r}{\kappa^2} C_3(h) y^{r_1-1} + \frac{2r}{\kappa^2} C_4(h) y^{r_2-1} + \frac{1}{r \beta}
$$ is either monotone or first increasing then decreasing. Combining with Step 1, we can verify the statement of the lemma. Indeed, the extreme point $\breve{y}$ of $\gamma(y)$ should satisfy the first order condition $\gamma'(\breve{y})=0$, i.e.
$$
C_3(h) (r_1-1) (y^*)^{r_1-r_2} + C_4(h) (r_2-1) = 0.
$$
Note that $C_3(h) < 0$, while $C_4(h)$ can be negative or positive. If $C_4(h) \leq 0$, there is no solution $\breve{y}$, hence $\gamma(y)$ is monotone.  If $C_4(h) >0$, there exists a unique real solution to the above equation that
$$
\breve{y}=\left( \frac{C_4(h)(1-r_2)}{C_3(h) (r_1-1)} \right)^{\frac{1}{r_1-r_2}},
$$
which might fall into the interval $[e^{(\lambda-1)\beta h}, e^{\lambda \beta h}]$. As we have that $C_3(h) < 0$ and
$$
\gamma'(y)=\frac{2r}{\kappa^2} y^{r_2-2} \big( C_3(h) (r_1-1) (y)^{r_1-r_2} + C_4(h) (r_2-1) \big),
$$
it follows that $\gamma'(y) \geq 0$ if and only if $y \leq \breve{y}$. Hence $\gamma(y)$ is increasing in $y$ before reaching $\breve{y}$ and is then decreasing in $y$ after $\breve{y}$.
\end{proof}

\begin{proof}[Corollary \ref{propasymp}]
As we consider the asymptotic behavior along the boundary $x_{\text{lavs}}(h)$, we first have
\begin{align*}
\lim_{h\to+\infty}\frac{c^*(x_{\text{lavs}}(h), h)}{x_{\text{lavs}}(h)}=\lim_{h\to+\infty}\frac{h}{x_{\text{lavs}}(h)}.
\end{align*}
Taking into account the explicit form of $x_{\text{lavs}}(h)$ in \eqref{x3-def}, we need to compute the two limits
\begin{align*}
\lim_{h\rightarrow+\infty} \frac{-C_5(h)r_1(1-\lambda)^{r_1-1}e^{(\lambda-1)(r_1-1)\beta h}}{h}=\lim_{h\rightarrow+\infty}\frac{\frac{-r_1(1-\lambda)^{r_1-1}(1-r_2)\kappa^2}{2(r_1-r_2)\beta r^2}(1-e^{(1-r_1)\beta h})}{h}=0
\end{align*}
and
\begin{align*}
&\lim_{h\rightarrow+\infty} \frac{-C_6(h)r_2(1-\lambda)^{r_2-1}e^{(\lambda-1)(r_2-1)\beta h}}{h}\\
=&\lim_{h\rightarrow+\infty}\frac{\frac{-r_2(1-\lambda)^{r_1-1}(r_2-1)\kappa^2}{2(r_1-r_2)\beta r^2} ( \frac{1-r_1}{1-r_2}-\frac{\lambda(1-r_1)}{\lambda(1-r_2)-(r_1-r_2)}e^{(1-r_1)\beta h}) }{h}=0.
\end{align*}
Therefore, we obtain that
$$
\lim_{h\to+\infty}\frac{c^*(x_{\text{lavs}}(h), h)}{x_{\text{lavs}}(h)}=r.
$$
Similarly, thanks to the explicit form of $\pi^{*}(x,h)$ in \eqref{invest:primal}, we need to compute two limits along $x_{\text{lavs}}(h)$ that
\begin{align*}
\lim_{h\rightarrow+\infty}\frac{2r}{\kappa^2}C_5(h)(1-\lambda)^{r_1-1}e^{(\lambda-1)\beta h (r_1-1)}&=\lim_{h\rightarrow+\infty}\frac{(1-\lambda)^{r_1-1}(1-r_2)}{(r_1-r_2)\beta r}(1-e^{(1-r_1)\beta h})\\
&=\frac{(1-\lambda)^{r_1-1}(1-r_2)}{(r_1-r_2)\beta r},
\end{align*}
and
\begin{align*}
&\lim_{h\rightarrow+\infty}\frac{2r}{\kappa^2}C_6(h)(1-\lambda)^{r_2-1}e^{(\lambda-1)(r_2-1)\beta h}\\
=&\lim_{h\rightarrow+\infty}\frac{(1-\lambda)^{r_1-1}(r_2-1)}{(r_1-r_2)\beta r}\left( \frac{1-r_1}{1-r_2}-\frac{\lambda(1-r_1)}{\lambda(1-r_2)-(r_1-r_2)}e^{(1-r_1)\beta h}\right)=\frac{(1-\lambda)^{r_1-1}(r_1-1)}{(r_1-r_2)\beta r}.
\end{align*}
Therefore, we conclude that
\begin{align*}
\lim_{h\to+\infty}\pi^*(x_{\text{lavs}}(h), h) &=\frac{\mu-r}{\sigma^2}\left(\frac{(1-\lambda)^{r_1-1}(1-r_2)}{(r_1-r_2)\beta r}+ \frac{(1-\lambda)^{r_1-1}(r_1-1)}{(r_1-r_2)\beta r}\right) \\
&=\frac{(\mu-r)(1-\lambda)^{r_1-1}}{r\beta\sigma^2}.
\end{align*}
\end{proof}

\subsection{Proofs of Theorem \ref{verthm} and Corollary \ref{value_function_primal} }\label{sec:verification}

\proof[Theorem \ref{verthm}]
The proof of the verification theorem boils down to show that the solution of the PDE indeed coincides with the value function. In other words, there exists $(\pi^*, c^*) \in \mathcal{A}(x)$ such that
$$
u(x, h) = \E_{}\left[\int_0^{\infty} e^{- r t} U({c_t^* - \lambda H_t^*}) dt\right].
$$

For any admissible strategy $(\pi,c) \in \mathcal{A}(x)$, similar to the standard proof of Lemma 1 in Arun \cite{Arun}, we have the following budget constraint:
\begin{equation*} 
\E \left[\int_0^{\infty} c_t M_t dt \right] \leq x.
\end{equation*}

Regarding $(\lambda,h)$ as fixed parameters, we consider the dual transform of $U$ with respect to $c$ in the constrained domain that
\begin{align*}
V(y,h) &:=  \sup_{0\leq c \leq h} (U(c-\lambda h ) - cy) \; = \\&\left\{
\begin{aligned}
& - \frac{1}{\beta} e^{\lambda \beta h}, & & \mbox{if } y \geq e^{\lambda \beta h}, \\
& -\frac{1}{\beta}y +  y ( \frac{1}{\beta} \ln y - \lambda h),  & & \mbox{if } e^{(\lambda-1)\beta h} < y < e^{\lambda \beta h}, \\
& - \frac{1}{\beta}e^{(\lambda-1) \beta h} -  hy,  & & \mbox{if } (1-\lambda)e^{(\lambda-1)\beta h}\leq  y \leq e^{(\lambda-1) \beta h}.
\end{aligned}
\right.
\end{align*}
We remark that when $\lambda=0$, $V(y,h)$ is independent of $h$. Moreover, $V(y,h)$ can be attained by the construction of the feedback function $c^{\dagger}(y,h)$ given in \eqref{feedbackcp}.

In what follows, we distinguish the two reference processes, namely $H_t: =h\lor \sup_{s \leq t} c_s$ and $H_t^{\dagger}(y): =h\lor \sup_{s \leq t} c^{\dagger}(Y_s(y),H_s^{\dagger}(y))$ that correspond to the reference process under an arbitrary consumption process $c_t$ and under the optimal consumption process $c^{\dagger}$ with an arbitrary $y>0$. Note that the (global) optimal reference process will be defined later by $H_t^*:= H_t^{\dagger}(y^*)$ with $y^*>0$ to be determined.
Let us now further introduce
\begin{align}\label{defhatH}
{\hat H_t(y)} := h \lor \Bigg( \frac{1}{(\lambda-1)\beta} \ln\left( \frac{1}{1-\lambda} \inf_{s \leq t} Y_s(y) \right) \Bigg),
\end{align}
where $Y_t(y) \; = \; y e^{r t} M_t$ is the discounted martingale measure density process.

For any admissible $(\pi,c)$ $\in$ $\mathcal{A}(x)$ and all $y > 0$, we have that
\begin{align}
\E_{} \left[\int_0^{\infty} e^{- r t} U({c_t - \lambda  H_t}) dt \right] =& \E_{}\left[\int_0^{\infty} e^{- r t} \big(U(c_t - \lambda  H_t) - Y_t(y) c_t \big) dt \right] \nonumber \\
&+ y \E_{} \left[\int_0^{\infty} c_t M_t dt \right]  \nonumber \\
\leq & \E_{} \left[\int_0^{\infty} e^{- r t} V\big(Y_t(y), H_t^{\dagger}(y)\big) dt \right] + yx \label{ineg} \\
=   & \E_{} \left[\int_0^{\infty} e^{- r t} V\big(Y_t(y), {\hat H_t(y)} \big) dt \right] + yx \nonumber \\
= & v (y, h) +y x. \nonumber
\end{align}
where the second line follows from Lemma \ref{lemma:IneqAttained}, the third line holds thanks to Lemma \ref{lemma:ReplaceHat} below, and the last line is consequent on Lemma \ref{lemma:DualVerification}. In addition, by Lemma \ref{lemma:IneqAttained}, the inequality becomes equality with the choice of
$c^*_t=c^{\dagger}(Y_t(y^*), H_t^{\dagger}(y^*))$, in which $y^*$ uniquely solves $\E [\int_0^{\infty}  c^{\dagger}(Y_t(y),H_t^{\dagger}(y)) M_t dt ]=x$ for the given $x>0$ and $h\geq 0$.

In conclusion, we arrive at
$$
\sup_{(\pi,c) \in \mathcal{A}(x)} \E_{} \Big[\int_0^{\infty} e^{- r t} U({c_t - \lambda  H_t}) dt \Big] = \inf_{y > 0} \big( v (y, h) +yx\big)= u(x, h),
$$
which completes the proof of the verification theorem.
\smartqed

\vspace{1mm}

We then proceed to prove some auxiliary results that have been used to support the previous proof of the main theorem. We shall use the following asymptotic results of the  coefficients
defined in  Proposition \ref{dual_value_function}.

\begin{rem}\label{remarkord}
Based on the explicit formulas in \eqref{C2h}-\eqref{C4h}, we note that as $h\to+\infty$, we have the asymptotics 
$$
\begin{aligned}
& C_2(h) = O \left( e^{(\lambda-1)(1-r_2)\beta h} \right) + O \left( e^{(\lambda(1-r_2)-(r_1-r_2))\beta h} \right) + O\left( e^{\lambda  (1-r_2)\beta h} \right),\\
& C_3(h) = O \left( e^{\lambda\beta h (1-r_1)}  \right), \\
& C_4(h) = O \left( e^{(\lambda-1)(1-r_2)\beta h} \right) + O \left( e^{(\lambda(1-r_2)-(r_1-r_2))\beta h} \right), \\
& C_5(h) = O  \left( e^{\lambda\beta h (1-r_1)}  \right) +  O \left( e^{(\lambda-1)(1-r_1)\beta h} \right), \\
& C_6(h) = O \left( e^{(\lambda-1)(1-r_2)\beta h} \right) + O \left( e^{(\lambda(1-r_2)-(r_1-r_2))\beta h} \right).
\end{aligned}
$$
\end{rem}

\vspace{1mm}

\begin{lem} \label{lemma:DualVerification}
$$
v(y, h) = \E_{} \left[\int_0^{\infty} e^{- r t} V\big(Y_t(y), \hat{H}_t(y)\big) dt \right].
$$
\end{lem}
\proof
Note that the martingale measure density process $M_t$ satisfies the equation
$$
d M_t = M_t (-r dt - \kappa d W_t).
$$
By \eqref{EDPtildeU2},  $v(y,h)$ satisfies the ODE
$$
\frac{\kappa^2}{2} y^2 v_{yy} - r v + V(y, h) = 0.
$$
By It\^o's formula, we have that
\begin{align} \label{eq:from_Ito}
d \Big(e^{- r t} v\big(Y_t(y), \hat{H}_t (y)\big) \Big) = &- e^{- r t} V\big(Y_t(y), \hat{H}_t (y)\big) dt - \kappa e^{-r t} v_y\big(Y_t(y), \hat{H}_t (y)\big)Y_t(y) d W_t  \nonumber \\
&+ e^{-r t} v_h\big(Y_t(y), \hat{H}_t (y)\big) d \hat{H}_t(y).
\end{align}
Let us define the stopping time
\begin{align*}
\tau_n:= \inf \left\{ t \geq 0 : Y_t(y) \geq n,~ 
\hat{H}_t(y) \geq \frac{1}{(\lambda-1)\beta} \ln \frac{1}{(1-\lambda)n} \right\}.
\end{align*}
By integrating \eqref{eq:from_Ito} from $0$ to $T \wedge \tau_n$ and taking expectation on both sides, we have that
\begin{align}\label{refito}
v(y, h) = \E_{}\left[\int_0^{T \wedge \tau_n} e^{-r t} V\big(Y_t(y), \hat{H}_t (y)\big) dt\right]
+ \E_{}\left[ e^{-r (T \wedge \tau_n)} v\big(Y_{T \wedge \tau_n}(y), \hat{H}_{T \wedge \tau_n} (y)\big)\right].
\end{align}
To wit, the integral term with respect to $d \hat{H}_t(y)$ vanishes as $\hat{H}_t(y)$ increases only when $c^*_t(y)=\hat{H}_t(y)$ and we have
$v_h(Y_t(y), \hat{H}_t (y))=0$ by the free boundary condition. In addition, the expectation of the integral of $d W_t$ vanishes as the local martingale
$$
\int_0^{T \wedge \tau_n} \kappa v_y\big(Y_t(y), \hat{H}_t (y)\big) y M_t d W_t
$$
becomes a true martingale thanks to the definition of $\tau_n$ and the fact that $v$ is of class $C^2$.

By passing to the limit as $n\rightarrow+\infty$, the first term in \eqref{refito} tends to $$\E_{}\left[\int_0^{T} e^{-r t} V\big(Y_t(y) \hat H_t (y)\big) dt\right]$$ by the monotone convergence theorem. This follows from the two facts below: first, we clearly have $V<0$ by its definition; second, by some calculations similar to (A.25) of Guasoni et al. \cite{GHR}, we can see that when $n$ tends to infinite, $\tau_n \geq T$ almost surely.
Moreover, the second term in \eqref{refito} can be written as
\begin{align}\label{twoexpt}
&\E_{} \left[e^{-r (T \wedge \tau_n)} v\big(Y_{T \wedge \tau_n}(y), \hat H_{T \wedge \tau_n} (y)\big)\right] \nonumber \\ 
= &\E_{}\left[e^{-r T} v\big(Y_{T}(y), \hat H_{T} (y)\big)\mathbf{1}_{\{T < \tau_n\}} \right]  + \E_{}\left[ e^{-r \tau_n} v\big(Y_{\tau_n}(y), \hat H_{\tau_n} (y)\big)  \mathbf{1}_{\{T \geq \tau_n\}}\right].
 \end{align}
As $n\rightarrow+\infty$, the first term in \eqref{twoexpt} clearly converges to $\E_{}[e^{-r T} v(Y_{T}(y), \hat H_{T} (y)) ]$, as when $n$ tends to infinite, $\tau_n > T$ almost surely. We will further show that the transversality condition holds in the sense that $\E_{}[e^{-r T} v(Y_{T}(y), \hat H_{T} (y)) ]$ converges to $0$ as $T\rightarrow+\infty$ in Lemma \ref{lemm:Transversality}.

We then claim that the second term in \eqref{twoexpt} also converges to $0$ as $n\rightarrow+\infty$. To see this, it follows by the definition of $\tau_n$ that $O ( Y_{\tau_n}(y)^{r_1} ) = O ( n^{r_1} )$, $O ( Y_{\tau_n}(y)^{r_2} ) = O ( n^{r_2} )$ and $O ( e^{(\lambda-1) \beta \hat H_{\tau_n}(y)} ) = O ( n )$. Using the expressions of $v(y, h)$ in different regions, we can analyze the orders  in terms of $n$ of $C_2(\hat H_{\tau_n}(y)) Y_{\tau_n}(y)^{r_2}$, $C_3(\hat H_{\tau_n}(y)) Y_{\tau_n}(y)^{r_1}+C_4(\hat H_{\tau_n}(y)) Y_{\tau_n}(y)^{r_2}$ and $C_5(\hat H_{\tau_n}(y)) Y_{\tau_n}(y)^{r_1}+C_6(\hat H_{\tau_n}(y)) Y_{\tau_n}(y)^{r_2}$. 

In view of  Remark \ref{remarkord}, we have for $n\to+\infty$ that
 $$
 \begin{aligned}
 & O \left( e^{(\lambda-1)(1-r_2)\beta \hat H_{\tau_n}(y)} \right) = O \left( n^{r_2-1} \right), O \left( e^{(\lambda(1-r_2)-(r_1-r_2))\beta \hat H_{\tau_n}(y)} \right) = O \left( n^{\frac{\lambda(1-r_2) - (r_1-r_2)}{1-\lambda}} \right), \\
& O \left( e^{\lambda(1-r_2)\beta \hat H_{\tau_n}(y)} \right) = O \left( n^{\frac{\lambda}{1-\lambda}(1-r_2)} \right), O \left( e^{\lambda(1-r_1)\beta \hat H_{\tau_n}(y)} \right) = O \left( n^{\frac{\lambda}{1-\lambda}(1-r_1)} \right), \\ 
& O \left( e^{(\lambda-1)(1-r_1)\beta \hat H_{\tau_n}(y)} \right) = O \left( n^{1-r_1} \right).
 \end{aligned}
 $$ 
Similar to the proof of (A.25) in \cite{GHR}, we can show that there exists some constant $C$ such that
$$
\E[ \mathbf{1}_{\{\tau_n \leq T\}} ] \leq n^{-2 \phi} (1 + y^{2 \phi}) e^{CT},
$$
for any $\phi \geq 1$. In particular, if we choose
$$
\phi:= \frac{1}{2} \max \{  r_2 + \frac{\lambda(1-r_2)}{1-\lambda}, \frac{\lambda}{1-\lambda}(r_1-1) + r_1 \} + 1, 
$$
we have that $\E[C_2(\hat H_{\tau_n}(y)) Y_{\tau_n}(y)^{r_2} \mathbf{1}_{\{T \geq \tau_n\}} ]$, $\E[C_3(\hat H_{\tau_n}(y)) Y_{\tau_n}(y)^{r_1}+C_4(\hat H_{\tau_n}(y)) Y_{\tau_n}(y)^{r_2} \mathbf{1}_{\{T \geq \tau_n\}}]$ and $\E[C_5(\hat H_{\tau_n}(y)) Y_{\tau_n}(y)^{r_1}+C_6(\hat H_{\tau_n}(y)) Y_{\tau_n}(y)^{r_2} \mathbf{1}_{\{T \geq \tau_n\}}]$ tend to $0$ as $n\to+\infty$.\ As a result, we obtain the desired claim that
$$
\lim_{ n\rightarrow+\infty}\E_{}\left[ e^{-r \tau_n} v\big(Y_{\tau_n}(y), \hat H_{\tau_n} (y)\big)  \mathbf{1}_{\{T > \tau_n\}}\right]=0.
$$
\smartqed

\begin{lem} \label{lemma:ReplaceHat}
For all $y >0$, we have $H_t^{\dagger}(y)= \hat H_t(y)$, $t\geq 0$, and hence
$$
\E_{} \left[\int_0^{\infty} e^{- r t} V\big(Y_t(y), H_t^{\dagger}(y)\big) dt \right]
=  \E_{} \left[\int_0^{\infty} e^{- r t} V\big(Y_t(y), \hat H_t(y) \big) dt \right] .
$$
\end{lem}
\proof The proof is similar to Lemma $A.1$ in \cite{GHR}. For the sake of completeness, we present the argument here. Suppose that $H_t^{\dagger}(y)$ is strictly increasing at $t$, then $H_t^{\dagger}(y)=c^{\dagger}(Y_t(y), H_t^{\dagger}(y))$, and the optimal consumption is $c^{\dagger}(Y_t(y), H_t^{\dagger}(y))$ $=$ $\frac{1}{(\lambda-1)\beta} \ln (\frac{1}{1-\lambda} Y_t(y)) $. 

Define
\begin{align*}
\mathcal{I}_t:= \left\{ s \leq t: H_t^{\dagger}(y)\ \text{is strictly increasing at}\ s \right\}.
\end{align*}
We can now derive that
\begin{align*}
 H_t^{\dagger}(y) &= h \vee \sup_{s \in \mathcal{I}_t} c^{\dagger}(Y_s(y), H_s^{\dagger}(y)) = h \vee \sup_{s \in \mathcal{I}_t} \frac{1}{(\lambda-1) \beta} \ln \left(\frac{1}{1-\lambda} Y_s(y)\right) \\
 & =  h \vee \sup_{s \in \mathcal{I}_t} \frac{1}{(\lambda-1) \beta} \ln \left(\frac{1}{1-\lambda} Y_s(y)\right) \vee  \sup_{s \notin \mathcal{I}_t} \frac{1}{(\lambda-1) \beta} \ln \left(\frac{1}{1-\lambda} Y_s(y)\right)   \\
 & = h \vee \sup_{s \leq t} \frac{1}{(\lambda-1)\beta} \ln \left(\frac{1}{1-\lambda} Y_s(y)\right) = \hat{H}_t (y).
\end{align*}

 In the second line on the above, we have used the following fact: for any $s \notin \mathcal{I}_t$, from the condition that  $Y_s(y) > (1-\lambda)e^{(\lambda-1)\beta H_s^{\dagger}(y)}$, we can obtain that  
 $$
 \frac{1}{(\lambda-1)\beta} \ln (\frac{1}{1-\lambda} Y_s(y)) < H_s^{\dagger}(y)\leq H_t^{\dagger}(y) = h \vee \sup_{s \in \mathcal{I}_t} \frac{1}{(\lambda-1) \beta} \ln \left(\frac{1}{1-\lambda} Y_s(y)\right) .
 $$ 
 Hence
 $$
 \sup_{s \notin \mathcal{I}_t} \frac{1}{(\lambda-1) \beta} \ln \left(\frac{1}{1-\lambda} Y_s(y)\right) \leq h \vee \sup_{s \in \mathcal{I}_t} \frac{1}{(\lambda-1) \beta} \ln \left(\frac{1}{1-\lambda} Y_s(y)\right).
 $$
\ \\
\begin{lem} \label{lemma:IneqAttained}
The inequality \eqref{ineg} holds, and it becomes an equality with the consumption control $c_t^*=c^{\dagger}(Y_t(y^*),\hat H_t(y^*))$, $t$ $\geq$ $0$, with $y^*=y^*(x,h)$ as the unique solution to $$\E \left[\int_0^{\infty} c^{\dagger}\big(Y_t(y^*),\hat H_t(y^*)\big) M_t dt \right]=x$$.
\end{lem}
\proof
Using the definition of $V$, for all $(\pi, c) \in \mathcal{A}(x)$, $U({c_t - \lambda  H_t})- Y_t(y) c_t \leq V(Y_t(y), {H_t})$. Moreover, for any fixed $y>0$, the inequality becomes an equality with the optimal feedback $c^{\dagger}(Y_t(y), H_t^{\dagger}(y))$. In other words, for any admissible $(c_t)_{0 \leq t \leq T}$,  we have that for all $t \in [0,T]$,
\begin{align*}
U({c_t - \lambda  H_t})- Y_t(y) c_t \leq& U\left({c^{\dagger}\big(Y_t(y), H_t^{\dagger}(y)\big) - \lambda  H_t^{\dagger}(y)}\right)- Y_t(y) c^{\dagger}\big(Y_t(y), H_t^{\dagger}(y)\big)\\
 = &V\big(Y_t(y), {H_t^{\dagger}(y)}\big).
\end{align*}
Multiplying both sides by $e^{- r t}$ and integrating from $0$ to $T$, we have that
$$
\int_0^{\infty} e^{- r t} \big(U(c_t - \lambda  H_t) - Y_t(y) c_t \big) dt \leq \int_0^{\infty} e^{- r t} V\big(Y_t(y), {H_t^{\dagger}(y)}\big) dt.
$$
To turn \eqref{ineg} into an equality, the consumption should take the optimal feedback form $c^{\dagger}$, and $\E [\int_0^{\infty} c^{\dagger}(Y_t(y^*),\hat H_t(y^*)) M_t dt ] = x$ should be valid with some $y^*>0$.  
 We now show the existence of such $y^*>0$. To this aim, we introduce
$$
c^{\dagger}(Y_t(y),\hat H_t(y)) : = \hat H_t(y)  F_t\big(y,Y_t(y)\big),
$$
where the function $F$ is defined as 
$$
F_t(y,z):=\mathbf{1}_{\{ (1-\lambda) e^{-(1-\lambda) \beta \hat H_t(y)} \leq z \leq e^{-(1-\lambda) \beta \hat H_t(y)} \}}
+ \left(\lambda-\frac{\mbox{ln} z}{\beta\hat{H}_t(y)} \right) \mathbf{1}_{ \{ e^{-(1-\lambda) \beta \hat H_t(y)} \leq z \leq e^{\lambda \beta \hat H_t(y)} \} }.
$$

In view of the definition of $\hat{H}_t(y)$ in \eqref{defhatH}, one can obtain that: (i) If $y \downarrow 0$, then $\hat H_t(y) \uparrow +\infty$ and $F_t(y,Y_t(y)) > 0$, which yields that $\E_{} [\int_0^{\infty} M_t  c^{\dagger}(Y_t(y),\hat H_t(y)) dt ] \uparrow +\infty$; (ii) If $y \uparrow +\infty$, then $\hat H_t(y) \downarrow h$ and $F_t(y,Y_t(y)) \downarrow 0$, which yields that
$\E_{} \int_0^{\infty} M_t  c^{\dagger}(Y_t(y),\hat H_t(y)) dt ] \downarrow 0$. 

The existence of $y^*$ satisfying $\E [\int_0^{\infty} c^{\dagger}(Y_t(y^*),\hat H_t(y^*)) M_t dt ] = x$ is a consequence of the asymptotic results (i) and (ii), given the fact that $\E_{} [\int_0^{\infty} M_t  c^{\dagger}(Y_t(y),\hat H_t(y)) dt ]$ is continuous in the variable $y$.

\smartqed
\ \\
We then prove the transversality condition, which is a key step in the proof of Lemma \ref{lemma:DualVerification}:
\begin{lem} \label{lemm:Transversality}
For all $y$ $>$ $0$, the following transversality condition holds:
$$
\lim_{T \rightarrow + \infty} \E_{}\left[ e^{-r T} v\big(Y_T(y), \hat H_{T} (y)\big)\right] = 0.
$$
\end{lem}

\proof
Let us first recall that
$$
\hat H_t(y) =  h \lor \Bigg( \frac{1}{(\lambda-1)\beta} \ln\left(\frac{1}{1-\lambda}  \inf_{s \leq t} Y_s(y)\right) \Bigg).
$$

From Proposition \ref{dual_value_function}, in the interval $e^{(\lambda-1) \beta h} < y < e^{\lambda \beta h}$, which corresponds to the case $0 < c_t < H_t$, we have
$$
v(y,h)=C_3(h) y^{r_1} + C_4(h) y^{r_2} -\frac{y}{r \beta}+\frac{y}{r \beta} \left(\ln y-\lambda\beta h+\frac{\kappa^2}{2r}\right).
$$

In the interval $y \geq e^{\lambda \beta h}$, which corresponds to the case $c_t=0$, we have
$$
v(y,h)=C_2(h) y^{r_2} - \frac{1}{r \beta} e^{\lambda \beta h} ,
$$

In the interval $(1-\lambda) e^{(\lambda-1) \beta h}  \leq y \leq e^{(\lambda-1) \beta h}$, we have
$$
v(y,h) = C_5(h) y^{r_1} + C_6(h) y^{r_2} -\frac{1}{r}hy-\frac{1}{r\beta} e^{(\lambda-1)\beta h} ,
$$

a) We first deal with the case $0<c_t< H_t$ and check the asymptotic behavior of the following expectation
\begin{align*}
\mathbb{E}\Bigg[&e^{-r T} \Bigg( C_3(\hat H_T(y)) (Y_T(y))^{r_1} + C_4(\hat H_T(y)) (Y_T(y))^{r_2}\\&-\frac{ Y_T(y)}{r \beta}+\frac{Y_T(y) }{r \beta} \Big(\ln Y_T(y)-\lambda \beta \hat H_T(y)+\frac{\kappa^2}{2r}\Big) \Bigg)\Bigg].
\end{align*}

We consider its asymptotic behavior term by term.

(i) Let us start by considering the asymptotic behavior of the third term $\mathbb{E}[-e^{- rT}\frac{Y_T(y)}{r\beta}]$ and the fourth term $\mathbb{E}[e^{- rT}\frac{Y_T(y)}{r\beta}(\ln Y_T(y)-\lambda\beta \hat{H}_T(y)+\frac{\kappa^2}{2r})]$. For the third term, it is easy to see that
\begin{align}\label{thirdconv}
\E_{}\left[y e^{-(r+\frac{\kappa^2}{2})T-\kappa W_T} \frac{1}{r \beta}\right]= \frac{y}{r \beta} e^{-(r+\frac{\kappa^2}{2})T} \E_{}\left[e^{- \kappa W_T}\right]=\frac{y}{r \beta} e^{-r T},
\end{align}
which converges to $0$ as $T \rightarrow +\infty$. For the fourth term, we have that
\begin{align*}
&\frac{y M_T}{r \beta}  \left(\ln Y_T(y)-\lambda \beta \hat H_T(y)+\frac{\kappa^2}{2r}\right) \\=& \frac{1}{r \beta} \Bigg( y M_T \left(r T + \ln y +\frac{\kappa^2}{2r}\right) + y M_T \ln M_T - y M_T \lambda \beta \hat H_T(y) \Bigg).
\end{align*}
Similarly to \eqref{thirdconv}, we can show that $\mathbb{E}[y M_T (r T + \ln y + \frac{\kappa^2}{2r})]$ converges to $0$ and 
\begin{align*}
\E_{} [ y M_T \ln M_T ] &= -y e^{-(r+\frac{\kappa^2}{2})T} \left( \E_{}\left[ \kappa W_T e^{-\kappa W_T} \right] + \left(r+\frac{\kappa^2}{2}\right) T \E_{}\left[e^{- \kappa W_T}\right]   \right)
\\&=-y e^{-rT} \left(r-\frac{\kappa^2}{2}\right) T,
\end{align*}
which also converges to $0$ as $T\rightarrow+\infty$. Furthermore, we can deduce that
\begin{align*}
\E_{} [ y M_T \hat H_T(y) ] &\leq \E[y M_T h] + \E \left[ y M_T \frac{1}{(\lambda-1) \beta} \ln \left( \frac{1}{1-\lambda} y \inf_{s \leq T} (e^{r s} M_s)\right) \right] \\
&= O \left( \E[ y M_T ] \right) + O \left( e^{-r T} \E\left[ e^{- \kappa W_T - \frac{1}{2} \kappa^2 T } \sup_{s \leq T} \left( \kappa W_s + \frac{1}{2} \kappa^2 s \right) \right] \right).
\end{align*}
The first term $O(\mathbb{E}[yM_T])$ clearly vanishes as $T\to+\infty$ by repeating similar computations as above for showing that $\mathbb{E}[yM_T\ln M_T] \to 0$. For the second term, we first note that
\begin{align*}
&\E [e^{-\kappa W_T} \sup_{s\leq T}W_s] \\=&\sqrt{\frac{T}{2\pi}} - e^{\frac{1}{2} \kappa^2 T} \kappa T \Phi(-\kappa \sqrt{T}) + e^{\frac{1}{2} \kappa^2 T} \frac{1}{2\kappa} \left( \Phi(\kappa \sqrt{T}) - \Phi (-\kappa \sqrt{T}) \right).
\end{align*}
Let us define the equivalent measure $\Q$ under which $W_t^{(\frac{\kappa}{2})}:=W_t + \frac{\kappa}{2}t$ is a Brownian motion, with the Randon-Nikodym derivative
$$
\frac{d \Q}{d \P} \Big|_{\mathcal{F}_t}:= \exp \left( -\frac{1}{2} \kappa W_t - \frac{1}{8} \kappa^2 t  \right).
$$
It follows by Girsanov's theorem that
\begin{align*}
& e^{-r T} \E\left[ e^{- \kappa W_T - \frac{1}{2} \kappa^2 T } \sup_{s \leq T} \left( \kappa W_s + \frac{1}{2} \kappa^2 s \right) \right]    \\
=& ~~ \kappa e^{-r T} \E^{\mathbb{Q}}\left[ e^{- \kappa W_T^{(\frac{\kappa}{2})} } \sup_{s \leq T} W_s^{(\frac{\kappa}{2})}  \exp \left( \frac{1}{2} \kappa W_t^{(\frac{\kappa}{2})} - \frac{1}{8} \kappa^2 t  \right) \right] \\
= &~~ \kappa e^{-rT} \left( \sqrt{\frac{T}{2\pi}} e^{-\frac{1}{8} \kappa^2 T} - \frac{1}{2} \kappa T \Phi(-\frac{1}{2} \kappa \sqrt{T}) + \frac{1}{\kappa} \left( \Phi(\frac{1}{2} \kappa \sqrt{T}) - \Phi (-\frac{1}{2} \kappa \sqrt{T}) \right) \right),
\end{align*}
which clearly vanishes when $T \rightarrow +\infty$.

\vspace{2mm}

(ii) Let us continue to consider the term with $C_3(h)$. In view of the constraint $Y_T(y) < e^{\lambda \beta \hat H_T}$, we have
$$
\lambda \beta \hat H_T (1-r_1) < (1-r_1) \ln \big( Y_T(y)\big)
$$
and it follows that
$$
\begin{aligned}
& \ \E_{}\left[ e^{-r T} C_3\big(\hat H_T(y)\big) \big(Y_T(y)\big)^{r_1} \right] \\
&= O\left( \E_{}\left[ e^{\lambda \beta \hat H_T(y) (1-r_1)} e^{-r T} \big(Y_T(y)\big)^{r_1} \right] \right) \\
&\leq O\left( \E_{}\left[   \big(Y_T(y)\big)^{1-r_1} e^{-r T} \big(Y_T(y)\big)^{r_1} \right]    \right) \\
&=O\left( \E_{}\left[   y M_T    \right]    \right) \; =O\left(  e^{-rT}   \right). \\
\end{aligned}
$$
It is thus verified that the term $\mathbb{E}[e^{-rT}C_3(\hat{H}_T(y))(Y_T(y))^{r_1}]$ converges to $0$ as $T\rightarrow+\infty$.

(iii) Now let us work with the term $C_4(\hat H_T(y)) e^{-r T} (Y_T(y))^{r_2}$. Remark \ref{remarkord} asserts that
$$
C_4(h) = O \left( e^{(\lambda-1)(1-r_2)\beta h} \right) + O \left( e^{(\lambda(1-r_2)-(r_1-r_2))\beta h} \right).
$$

Let us define the set
\begin{align*}
A:=\left\{\frac{1}{\lambda-1} \ln\left(\frac{1}{1-\lambda} y \inf_{s \leq T} (e^{r s} M_s) \right) \geq  h \right\}=\left\{\inf_{s\leq T} (e^{r s} M_s) \leq(1-\lambda) e^{(\lambda-1)h}\frac{1}{y} \right\},
\end{align*}
and two auxiliary random variables
\begin{align*}
G^1:= &\left(\frac{y}{1-\lambda}\right)^{(1-r_2)\beta} \inf_{s \leq T} \left(e^{r s} M_s \right)^{(1-r_2)\beta},\\
G^2:=  &\left(\frac{y}{1-\lambda}\right)^{\frac{\lambda(1-r_2)-(r_1-r_2)}{\lambda-1}\beta} \inf_{s \leq T} \left(e^{r s} M_s \right)^{\frac{\lambda(1-r_2)-(r_1-r_2)}{\lambda-1}\beta}.
\end{align*}

Using the formula of $\hat{H}_t$ defined in \eqref{defhatH}, we have that
$$
\begin{aligned}
&\E\left[ e^{-r T} C_4\big(\hat H_T(y)\big) \big(Y_T(y)\big)^{r_2} \right] \\
=& O\left( \E_{}\left[ e^{(\lambda-1) (1-r_2) \beta \hat H_T(y) } e^{-r T} (Y_T(y))^{r_2} + e^{(\lambda(1-r_2)-(r_1-r_2)) \beta \hat H_T(y)}  e^{-r T} (Y_T(y))^{r_2} \right] \right) \\
=& O \Big(  \E_{} \Big[ G^1 (e^{r T} M_T)^{r_2} e^{-r T} \mathbf{1}_{A} \Big] \Big) \vee O \Big( \E_{} \Big[ G^2 (e^{r T} M_T)^{r_2} e^{-r T} \mathbf{1}_{A} \Big]  \Big) \\& \vee O \Big(  \E_{} \Big[ e^{-r T} \big(Y_T(y)\big)^{r_2} \mathbf{1}_{A^c}  \Big] \Big)\\
=& O(\Upsilon_1(T)) \vee O(\Upsilon_2(T))  \vee O \Big(  \E_{} \Big[ e^{-r T} \big(Y_T(y)\big)^{r_2} \mathbf{1}_{A^c}  \Big] \Big)  ,
\end{aligned}
$$
in which we define
\begin{align*}
\Upsilon_1(T):=& \E_{} \left[ \left(\frac{y}{1-\lambda}\right)^{(1-r_2)\beta} \left( \inf_{s \leq T} \left(e^{r s} M_s \right) \right)^{(1-r_2)\beta} (e^{r T} M_T)^{r_2} e^{-r T} \mathbf{1}_{A} \right],  \\
\Upsilon_2(T):=& \E_{} \left[ \left(\frac{y}{1-\lambda}\right)^{\frac{\lambda(1-r_2)-(r_1-r_2)}{\lambda-1}\beta} \left( \inf_{s \leq T} \left(e^{r s} M_s \right) \right)^{\frac{\lambda(1-r_2)-(r_1-r_2)}{\lambda-1}\beta} (e^{r T} M_T)^{r_2} e^{-r T} \mathbf{1}_{A} \right],
\end{align*}
and the third term $O(\mathbb{E}[e^{-rT}(Y_T(y))^{r_2}\mathbb{1}_{A^c}])$ comes from the level h in the definition of $\hat H_t(y)$ with $A^c$ being the complementary set of $A$. 

We then proceed to show that {all three terms $\Upsilon_1(T)$ and $\Upsilon_2(T)$ and $\E_{} [ e^{-r T} (Y_T(y))^{r_2} \mathbf{1}_{A^c}  ]$ }converge to $0$ as $T\rightarrow+\infty$.

By setting $$a_1=-\kappa r_2, \ b_1=-\kappa (1-r_2)\beta, \ b_2=-\kappa \frac{\lambda(1-r_2)-(r_1-r_2)}{\lambda-1}\beta, \ \zeta=\frac{\kappa}{2},$$
we can use \cite[Corollary A.7]{GHR} to obtain 
\begin{align}\label{eq:auxi_bound_1}
&\lim_{T \rightarrow \infty}\frac{1}{T}\log \Upsilon_1(T) \nonumber  \\
=&\lim_{T \rightarrow \infty}\frac{1}{T}\log \left(  \E_{} \left[ \left(\frac{y}{1-\lambda}\right)^{(1-r_2)\beta} \left( \inf_{s \leq T} \left(e^{r s} M_s \right) \right)^{(1-r_2)\beta} (e^{r T} M_T)^{r_2} e^{-r T} \mathbf{1}_{A} \right]  \right) \nonumber \\
\leq& \max \left\{ \frac{a_1(a_1+2\zeta)}{2}-r, \frac{(a_1+b_1)(a_1+b_1+2\zeta)}{2}-r, -\frac{\zeta^2}{2}-r \right\}.
\end{align}
Similarly, we have that
\begin{align}\label{eq:auxi_bound_2}
&\lim_{T \rightarrow \infty}\frac{1}{T}\log \Upsilon_2(T)\nonumber \\
=&\lim_{T \rightarrow \infty}\frac{1}{T}\log \Bigg(  \E_{} \Bigg[ \left(\frac{y}{1-\lambda}\right)^{\frac{\lambda(1-r_2)-(r_1-r_2)}{\lambda-1}\beta} \nonumber \\& \left( \inf_{s \leq T} \left(e^{r s} M_s \right) \right)^{\frac{\lambda(1-r_2)-(r_1-r_2)}{\lambda-1}\beta} (e^{r T} M_T)^{r_2} e^{-r T} \mathbf{1}_{A} \Bigg]  \Bigg) \nonumber \\
\leq& \max \left\{ \frac{a_1(a_1+2\zeta)}{2}-r, \frac{(a_1+b_2)(a_1+b_2+2\zeta)}{2}-r, -\frac{\zeta^2}{2}-r \right\}.
\end{align}

We now show that the above bounds in \eqref{eq:auxi_bound_1} and \eqref{eq:auxi_bound_2} are either negative or not attainable. First, there is the same third bound $-\frac{\zeta^2}{2}-r<0$ in both \eqref{eq:auxi_bound_1} and \eqref{eq:auxi_bound_2}. For the same first bound $\frac{a_1(a_1+2\zeta)}{2}-r$ in \eqref{eq:auxi_bound_1} and \eqref{eq:auxi_bound_2}, direct calculations lead to
$$
\frac{a_1(a_1+2\zeta)}{2} -r = -\frac{1}{2}\kappa^2 r_1 (1-r_1) -r = 0.
$$
However, this zero bound can never be reached, as by \cite[Lemma A.5 and Corollary A.7]{GHR} the corresponding condition of attaining the bound $\frac{a_1(a_1+2\zeta)}{2}-r$ is $a_1 +\zeta < 0$, but we instead have that
$$
a_1 +\zeta = -\kappa r_2 + \frac{\kappa}{2}
= \frac{1}{\kappa} \sqrt{(-\frac{\kappa^2}{2})^2+2 r \kappa^2} > 0.
$$

We next claim that the second upper bound $\frac{(a_1+b_1)(a_1+b_1+2 \zeta)}{2}-r$ in \eqref{eq:auxi_bound_1} is strictly negative. From  \cite[Corollary A.7]{GHR}, this bound is attained if and only if $a_1+b_1+\zeta>0$ and $2 a_1 + b_1 + 2 \zeta>0$. Noting that $\kappa>0$, we have the equivalence
$$
\begin{aligned}
 2 a_1 + b_1 + 2 \zeta>0  
\Longleftrightarrow & ~ - 2 \kappa r_2 - \kappa(1-r_2) \beta + \kappa >0  \\
\Longleftrightarrow & ~ \beta < \frac{1-2r_2}{1-r_2}. \ \
\end{aligned}
$$

Therefore, under the condition $2 a_1 + b_1 + 2 \zeta>0$, i.e. $\beta < \frac{1-2r_2}{1-r_2}$, the upper bound $\frac{(a_1+b_1)(a_1+b_1+2 \zeta)}{2}-r$ must be negative because
$$
\begin{aligned}
& (a_1+b_1)(a_1+b_1+2 \zeta) - 2r  \\
&= \kappa^2 \big(r_2 + (1-r_2)\beta\big) \big(r_2 + (1-r_2) \beta - 1\big) - 2r \\
&= \kappa^2 \big( r_2^2 + (1-r_2)^2 \beta^2 + 2 r_2 (1-r_2) \beta - r_2 - (1-r_2) \beta   \big) - 2r \\
&= \kappa^2 \big( (1-r_2)^2 \beta^2 + 2 r_2 (1-r_2) \beta - (1-r_2) \beta   \big) \\
&=  \kappa^2 (1-r_2) \beta \big( (1-r_2)\beta + 2 r_2 -1 \big) < 0.\ \
\end{aligned}
$$

Finally, we claim that the second upper bound $\frac{(a_1+b_2)(a_1+b_2+2\zeta)}{2}-r$ in \eqref{eq:auxi_bound_2} is also strictly negative. Once again from \cite[Corollary A.7]{GHR}, this upper bound is attained if and only if $a_1+b_2+\zeta>0$ and $2 a_1 + b_2 + 2 \zeta>0$. As $\kappa>0$, we obtain the equivalence
$$
\begin{aligned}
 2 a_1 + b_2 + 2 \zeta>0  
\Longleftrightarrow & ~ - 2 \kappa r_2 -\kappa \frac{\lambda(1-r_2)-(r_1-r_2)}{\lambda-1}\beta + \kappa >0  \\
\Longleftrightarrow & ~ \beta < \frac{1-2r_2}{\lambda^*}, \ \
\end{aligned}
$$
where we define $\lambda^* :=  \frac{(r_1-r_2) - \lambda (1-r_2)}{1-\lambda}$. Hence, when the bound $\frac{(a_1+b_2)(a_1+b_2+2\zeta)}{2}-r$ is attained, the condition $2 a_1 + b_2 + 2 \zeta>0$, i.e. $\beta < \frac{1-2r_2}{\lambda^*}$, guarantees that this bound must be negative because
$$
\begin{aligned}
 (a_1+b_2)(a_1+b_2+2 \zeta) - 2r 
&= \kappa^2 (r_2 + \lambda^*\beta) (r_2 + \lambda^* \beta - 1) - 2r \\
&= \kappa^2 ( r_2^2 + \lambda^{*2} \beta^2 + 2 r_2 \lambda^* \beta - r_2 - \lambda^* \beta  ) - 2r \\
&= \kappa^2 ( \lambda^{*2} \beta^2 + 2 r_2 \lambda^* \beta - \lambda^* \beta  )  \\
&=  \kappa^2 \lambda^* \beta \left( \lambda^* \beta +2 r_2 - 1 \right) < 0.
\end{aligned}
$$

Putting everything together, we conclude that $\Upsilon_1(T)\rightarrow 0$ and $\Upsilon_2(T)\rightarrow 0$ as $T\rightarrow +\infty$.

The last term $\E_{} [ e^{-r T} (Y_T(y))^{r_2} \mathbf{1}_{A^c}  ]$ converges to $0$ as $T \rightarrow +\infty$ by Lemma \ref{lemma:BrownianMax} with $a= - \kappa r_2, b=0, \eta=\frac{r}{\kappa r_2} + \frac{1}{2} \kappa$.

\vspace{2mm}

b) Let us now deal with the case $c_t=0$. In this case, we need to calculate the order of $e^{-rT} C_2(\hat H_T(y)) (Y_T(y))^{r_2}- e^{-rT} \frac{1}{r \beta} e^{\lambda \beta \hat H_T(y)} $  when $T \rightarrow +\infty$. From Remark \ref{remarkord}, we recall that as $h \rightarrow +\infty$, $C_2$ has the order 
$$
O ( e^{(\lambda-1)(1-r_2)\beta h} ) + O ( e^{(\lambda(1-r_2)-(r_1-r_2))\beta h} ) + O( e^{\lambda  (1-r_2)\beta h} ).
$$
 As the first two terms $O ( e^{(\lambda-1)(1-r_2)\beta h} )$ and $O ( e^{(\lambda(1-r_2)-(r_1-r_2))\beta h} )$ are identical to the asymptotic expression of $C_4$ analysed in the previous case when $0 < c_t < H_t$, we only need to consider here $O( e^{\lambda \beta  h (1-r_2)} )$ and hence study the limit behaviour of $e^{- r T} e^{\lambda \beta \hat{H}_T(y) (1-r_2)} (e^{r T} y M_T)^{r_2}$ for $T \rightarrow +\infty$. Due to the condition $e^{\lambda \beta \hat{H}_T(y)} < e^{r T} y M_T$, we have
$$
e^{\lambda \beta \hat{H}_T(y) (1-r_2)} < (e^{r T} y M_T)^{1-r_2}.
$$
It follows that
$$
e^{- r T} e^{\lambda \beta \hat{H}_T(y) (1-r_2)} (e^{r T} y M_T)^{r_2} < y M_T.
$$
The term $e^{-rT} \frac{1}{r \beta} e^{\lambda \beta \hat H_T(y)}$ is also bounded by $\frac{1}{r \beta} y M_T$ using condition $e^{\lambda \beta \hat{H}_T(y)} < e^{r T} y M_T$. In the previous case $0 < c_t < H_t$,  it has been shown that $\E[ e^{r T} y M_T ]$ converges to $0$ as $T \rightarrow +\infty$, which verifies the claim in this case.

c) We now turn to the proof of the case $c_t=H_t$. Similar as before, we need to calculate the order of $$e^{-rT} \Big( C_5(\hat H_T(y)) (Y_T(y))^{r_1} + C_6(\hat H_T(y)) (Y_T(y))^{r_2} - \frac{1}{r} \hat H_T(y) Y_T(y) -\frac{1}{r\beta} e^{(\lambda-1)\beta \hat H_T(y)} \Big)$$ 
when $T \rightarrow +\infty$. By Remark \ref{remarkord}, when $h \rightarrow +\infty$, we have $$C_5(h) = O  \left( e^{\lambda\beta h (1-r_1)}  \right) +  O \left( e^{(\lambda-1)(1-r_1)\beta h} \right)$$ and $$C_6(h) = O \left( e^{(\lambda-1)(1-r_2)\beta h} \right) + O \left( e^{(\lambda(1-r_2)-(r_1-r_2))\beta h} \right).$$ 
Firstly, as $C_6(h)$ and $C_4(h)$ have the same asymptotic expressions, we can conclude that $\E[ e^{-rT} C_6(\hat H_T(y)) (Y_T(y))^{r_2} ]$ converges to $0$ as $T \rightarrow +\infty$, thanks to the asymptotic result  $\lim_{T \rightarrow +\infty}\E[ e^{-rT} C_4(\hat H_T(y)) (Y_T(y))^{r_2} ]=0$ in step 3 when $0 < c_t < H_t$. For the asymptotic form of $C_5(h)$, note that its first term $O  \left( e^{\lambda\beta h (1-r_1)}  \right)$ coincides with the asymptotic expression of $C_3(h)$ in Remark \ref{remarkord}. We can see from step 2 in the case $0 < c_t < H_t$ that $\E[ e^{-rT} e^{\lambda\beta \hat H_T(y) (1-r_1)} (Y_T(y))^{r_1}  ]$ converges to $0$ as $T \rightarrow +\infty$.
 For the second term $O \left( e^{(\lambda-1)(1-r_1)\beta h} \right)$, thanks to the condition $e^{(\lambda-1) \beta \hat{H}_T(y)} \leq Y_T(y)$, we hence have 
$$
e^{(\lambda-1) (1-r_1) \beta \hat{H}_T(y)} \leq (Y_T(y))^{1-r_1}.
$$
Following the same computations as in step 2 of the case $0 < c_t < H_t$, the desired result holds that $\E [ e^{-rT} C_5(\hat H_T(y)) (Y_T(y))^{r_1}  ]$ converges to $0$ as $T \rightarrow +\infty$ . The term $e^{-rT} \frac{1}{r\beta} e^{(\lambda-1)\beta \hat H_T(y)}$ is first bounded due to $e^{(\lambda-1) \beta \hat{H}_T(y)} < e^{r T} y M_T$. Using similar arguments as for the case $c_t=0$, we can obtain its convergence result that $\E[ e^{-rT} \frac{1}{r\beta} e^{(\lambda-1)\beta \hat H_T(y)} ]$ converges to $0$ as $T \rightarrow +\infty$. The last term $\frac{1}{r} \hat H_T(y) Y_T(y)$ term has already been handled in the proof for the case $0 < c_t < H_t$, which eventually completes the whole proof.
\smartqed

\vspace{2mm}

The following result has been used in the previous proof, which is essentially the same to Corollary A.7 of \cite{GHR}. We present it here for the completeness.
\begin{lem} \label{lemma:BrownianMax}
Let $B_t^{(\zeta)}=B_t+\zeta t$, where $B$ is a standard Brownian motion, and let $\overline{B}_t^{(\zeta)} $ be the running maximum of $B_t^{(\zeta)}$. Then for any constant $a,b,k$ with $2a+b+2 \zeta \neq 0$, $k \geq 0$, we have
$$
\begin{aligned}
&  \E \left[  e^{a B^{(\zeta)}_{T} + b  \overline{B}^{(\zeta)}_{T} } \mathbf{1}_{\left\{  \overline{B}^{(\zeta)}_{T}  \leq k \right\}}  \right]  \\
=& ~ \frac{2(a+b+c)}{2a +b +\zeta} \exp\left( \frac{(a+b)(a+b+2\zeta)}{2}T \right)\\&\times \left(  \Phi\left( (a+b+\zeta)\sqrt{T}\right) - \Phi\left( (a+b+\zeta)\sqrt{T} -\frac{k}{\sqrt{T}} \right) \right)\\
&+\frac{2(a+\zeta)}{2a+b+2 \zeta} \Bigg( \exp \left( \frac{a(a+2\zeta)}{2}T \right) \Phi\left( -(a+\zeta)\sqrt{T} \right)\\
&-\exp \left( (2a+b+2\zeta)k + \frac{a(a+2\zeta)}{2}T \right) \Phi\left( -(a+\zeta)\sqrt{T} -\frac{k}{\sqrt{T}} \right) \Bigg).
\end{aligned}
$$
In particular, we have that
$$
\lim_{T \rightarrow +\infty} \E \left[  e^{a B^{(\zeta)}_{T} + b \overline{B}^{(\zeta)}_{T} } \mathbf{1}_{\left\{  \overline{B}^{(\zeta)}_{T}  \leq k \right\}}  \right]  = 0.
$$
\end{lem}

%

\ \\

At last, to prove Corollary \ref{value_function_primal}, it is sufficient to prove the existence of the unique strong solution to the SDE \eqref{wealthSDE} for $X_t^*$ under optimal controls. First, we need to establish the following results concerning the regularity of the feedback functions $c^*(x,h)$ and $\pi^*(x,h)$.

By the definition of $g$ in \eqref{sol_for_x} and the fact that $f(\cdot, h)$ is the inverse of $g(\cdot, h)$, we have the following results of the function $f$.
\begin{lem} \label{f_regularity}
The function $f$ is $C^1$ within each of the three subsets of $\mathbb{R}_+^2$: $x \leq x_{\text{zero}}(h)$, $x_{\text{zero}}(h) < x < x_{\text{aggr}}(h)$ and $x_{\text{aggr}}(h) \leq x \leq x_{\text{lavs}}(h)$, and it is continuous at the points $x=x_{\text{aggr}}(h)$ and $x=x_{\text{lavs}}(h)$. Moreover, we have:
\begin{align}  \label{f_derivative_x}
&f_x(x,h) = \frac{1}{g_y(f,h)} \nonumber \\
=& \left\{
\begin{aligned}
&\Big(-C_2(h) r_2 (r_2-1)  \big(f_1(x,h)\big)^{r_2-2} \Big)^{-1} , & & \mbox{if } x \leq x_{\text{zero}}(h), \\
&\Bigg(-C_3(h) r_1 (r_1-1) \big(f_2(x,h)\big)^{r_1-2} & & \\
&\ \ \ - C_4(h) r_2 (r_2-1) \big(f_2(x,h)\big)^{r_2-2} -\frac{1}{r\beta f_2(x,h) } \Bigg)^{-1} , & & \mbox{if } x_{\text{zero}}(h) < x < x_{\text{aggr}}(h), \\
&\Bigg(-C_5(h) r_1(r_1-1) \big(f_3(x,h)\big)^{r_1-2}\\& - C_6(h) r_2 (r_2-1) \big(f_3(x,h)\big)^{r_2-2} + \frac{1}{r}h \Bigg)^{-1}, & & \mbox{if } x_{\text{aggr}}(h) \leq x \leq x_{\text{lavs}}(h),\\
\end{aligned}
\right.
\end{align}
and
\begin{equation}\label{f_derivative_z}
f_h(x,h)= -g_h\big(f(x,h),h\big) \cdot f_x(x,h).
\end{equation}
\end{lem}
\proof
The proof of the lemma is similar to lemma 6.1 of Elie and Touzi \cite{ElieTouzi}. As the inverse of $g$, the function $f$ satisfies 
\begin{equation*} 
g\big(f(x,h),h\big)=x, ~~~\mathrm{for}~(x,h) \in \mathbb{R}_+^2.
\end{equation*}
From the definition of $g$ in \eqref{sol_for_x}, we know that for fixed $h$, the map $x \mapsto g(x,h)$ is $C^1$ and decreasing. By the inverse function theorem, the map $x \mapsto f(x,h)$ is also $C^1$ and decreasing, for any $h > 0$.

Using the expression of $v$ in Proposition \ref{dual_value_function} and the definition of $g$ in \eqref{sol_for_x}, one can directly calculate the partial derivative of $g$ with respect to its first argument, and then get \eqref{f_derivative_x}. Similarly, we can calculate the partial derivative $g_h$ explicitly.
As $g_h$ is clearly a continuous function in each of the closed intervals $\{0 \leq x \leq x_{\text{zero}}(h)\}, \{x_{\text{zero}}(h) \leq x \leq x_{\text{aggr}}(h)\}, \{x_{\text{aggr}}(h) \leq x \leq x_{\text{lavs}}(h)\}$, it is bounded, 
i.e. $\exists$ a constant $\alpha > 0$, such that $g_h(x,h) \leq \alpha$, for all $(x,h) \in \mathbb{R}^2_+$. Now in order to prove that $f$ is $C^1$ within all three intervals $\{x \leq x_{\text{zero}}(h)\}$, $\{x_{\text{zero}}(h) < x < x_{\text{aggr}}(h) \}$ and $\{ x_{\text{aggr}}(h) \leq x \leq x_{\text{lavs}}(h) \}$, we can verify that $f$ is differentiable in each variable with continuous partial derivative.

First, let us prove that $f$ belongs to  $C^0$ in the three regions of $(x,h) \in \mathbb{R}_+^2$: $\{ 0 \leq x \leq x_{\text{zero}}(h)\}$, $\{ x_{\text{zero}}(h) \leq x \leq x_{\text{aggr}}(h) \}$, $\{ x_{\text{aggr}}(h) \leq x \leq x_{\text{lavs}}(h) \}$, which implies that $f_x \in C^0$ in each of the three regions (as $f_x(x,h)$ is a differentiable function of $f(x,h)$). Indeed, for a pair $(x,h)$ belonging to one of the intervals and a $l_2$ small enough, we have that
$$
\begin{aligned}
g\big(f(x, h+l_2), h\big) -x  &= g\big(f(x, h+l_2), h\big) -  g\big(f(x, h+l_2), h+l_2\big) \\
& \leq \alpha l_2 \longrightarrow 0, ~~~~~~~\mbox{as }  l_2 \rightarrow 0.
\end{aligned}
$$
Now using the continuity of $f(\cdot, h)$, we obtain
$$
f(x,h+l_2) - f(x,h) = f\Big(g\big(f(x, h+l_2),h\big),h\Big) - f(x,h) \longrightarrow 0, ~~~\mbox{as }  l_2 \rightarrow 0.
$$
Finally, for sufficiently small $l_1$, we have that
$$
f(x+l_1, h+l_2) - f(x,h) = f_x(x_l, h+l_2) l_1 + f(x, h+l_2) -f(x,h),
$$
which will tend to $0$ when $l_1, l_2$ tend to $0$, and this shows that $f$ is continuous at an arbitrary point $(x,h)$.

Secondly, let us show that $f$ is differentiable with respect to $h$ with continuous partial derivatives. Let the pair $(x,h)$ in a certain interval and $l$ small enough such that $(x, h+l)$ is in the same interval. We have that
$$
\begin{aligned}
\frac{1}{l} \left( f(x, h+l) - f(x,h) \right)  &= \frac{1}{l} \Big( f(x, h+l) - f\big(g(f(x,h),h+l),h+l\big) \Big)  \\
&= f_x(x_l, h+l) \frac{1}{l}\Big( g\big(f(x,h),h\big) - g\big(f(x,h), h+l\big) \Big),
\end{aligned}
$$
for some $x_l \in [x, x+ g(f(x,h),h+l)]$. As $f_x \in C^0$ and $g_h(f(x,h),\cdot)$ is continuous, we obtain
$$
\lim_{l \rightarrow 0} \frac{1}{l}  \Big( f(x, h+l) - f(x,h) \Big) = - f_x(x,h) g_h\big(f(x,h),h\big),
$$
which gives \eqref{f_derivative_z}. Then, the continuity of $f_h$ follows from \eqref{f_derivative_z} and the continuity of $f$.\smartqed

The next result guarantees the existence of the strong solution in Proposition \ref{prop:existence_strong} below. Its proof is standard and lengthy, which will be reported in Appendix  \ref{appB}.

\begin{lem} \label{Lipschitz_c_pi}
The functions $c^*$ is locally Lipschitz on  $\mathcal{C}$, and the function $\pi^*$ is Lipschitz on $\mathcal{C}$.
\end{lem}

We are now ready to verify that the SDE \eqref{wealthSDE} has a unique strong solution. 
\begin{prop} \label{prop:existence_strong}
The SDE \eqref{wealthSDE} has a unique strong solution $(X_t^*, H_t^*)$ for any initial condition $(x,h) \in \mathcal{C}$.
\end{prop}
\proof
\textit{Step 1}. First, we show that the stochastic differential equation
\begin{align}\label{wealthSDE_auxilliary}
d \bar{X}_t =  r \bar{X}_t dt+ \pi^*(\bar{X}_t,\bar{H}_t)(\mu -r)dt+ \pi^*(\bar{X}_t,\bar{H}_t)\sigma dW_t
\end{align}
has a unique strong solution. To see this, let us introduce the functionals
$$
G_1(t, x(t), h(t)) : =  r x(t) + \pi^*(x(t), h(t)) (\mu -r),
$$
and
$$
G_2(t, x(t), h(t)) := \pi^*(x(t), h(t)).
$$
By Lemma \ref{Lipschitz_c_pi}, we can obtain that both $G_1$ and $G_2$ are Lipschitz functions. This justifies the existence of strong solution for  the SDE \eqref{wealthSDE_auxilliary}.

\textit{Step 2}. Let us consider equation \eqref{wealthSDE}. Since  $c^*$ is locally Lipschitz on $\mathcal{C}$, a similar argument as the previous step implies the local existence and uniqueness of the stochastic differential equation \eqref{wealthSDE}. Using the fact that $c^* > 0$, it follows that $0 \leq X_t^* \leq \bar{X}_t$, hence there is no explosion of the local solution.

\smartqed


\appendix\normalsize

\section{The extreme case when $\lambda=1$}\label{appA}
We present here some computational results when $\lambda=1$. Solving the HJB equation essentially follows the same arguments when $0<\lambda<1$. However, the effective domain $\mathcal{C}$ defined in \eqref{primedomain} needs to be modified to
\begin{align}\label{C-2}
\mathcal{C}:=\left\{(x,h)\in\mathbb{R}_+\times \mathbb{R}_+: u_x(x,h)\geq 0\right\}.
\end{align}
Equivalently, $\mathcal{C}=\mathcal{R}_1\cup\mathcal{R}_2\cup\mathcal{R}_3=\mathbb{R}_+^2$, where $\mathcal{R}_1$, $\mathcal{R}_2$ are defined in the same way as in Section \ref{subsec:Heuristic} for $0<\lambda<1$.

In particular, we now consider $\mathcal{R}_3=\left\{(x,h)\in\mathbb{R}_+\times \mathbb{R}_+: 0\leq u_x(x,h)\leq 1\right\}$ and note that the previous auxiliary singular control 
\begin{align*}
\hat{c}(x)=\frac{1}{\beta(\lambda-1)}\ln\left(\frac{u_x}{1-\lambda}\right)
\end{align*}
for the case $0<\lambda<1$ is no longer well defined if we have $\lambda=1$.

In fact, in the extreme case $\lambda=1$, there is no need to consider the singular optimal consumption that exceeds the previous maximum level $h$. In the whole region $\mathcal{R}_3$, the optimal consumption is no longer unique, but one feedback optimal consumption is to constantly consume the initial level $H_0^*=h$ such that $c^*(x,h)\leq h$ can be guaranteed for any $x\geq 0$. That $H_t^*$ will never increase for $\lambda=1$ results from the formulation $U(c_t-H_t)$ where the utility is defined on the difference. For the case $0<\lambda<1$, the utility $U(c_t-\lambda H_t)$ allows the investor to gain positive outperformance $c_t-\lambda H_t>0$ if he chooses a large $c_t$ to increase $H_t$. On the other hand, for the case $\lambda=1$, the investor can only obtain $0=c_t-H_t$ by choosing to consume more than the past maximum. However, the investor can also easily achieve the same goal of zero difference $c_t-H_t$ by following the previously attained maximum level without creating any new record. Therefore, to achieve the largest gap $c_t-H_t=0$, one optimal way is to sit on the previous consumption peak and the investor has no incentives to increase the reference process $H_t$ at any time. Consequently, we shall only adopt the feedback control $c^*(x,h)=h$ in the region $\mathcal{R}_3$.

Based on the observations above, if the wealth $x$ is larger than or equal to the subsistence level $x\geq x^*:=\frac{h}{r}$, the investor can always choose to invest zero amount $\pi^*_t\equiv 0$ in the risky asset and save the initial wealth $x^*$ in the bank account such that the interest rate can support the constant consumption at the initial reference level $c^*_t=H_0=h$, $t\geq 0$. That is, we have $c^*_t-H^*_t=0$ for $t\geq 0$. As a consequence, the value function defined in \eqref{primalvalue} attains its maximum value $u(x,h)=-\frac{1}{r\beta}$ for $x\geq \frac{h}{r}$. The primal value function $u(x,h)$ for $\lambda=1$ is no longer strictly concave and $u(x,h)$ remains constant (and $u_x(x,h)=0$) for $x\geq\frac{h}{r}$, which differs substantially from the case $0<\lambda<1$. Therefore, we have the asymptotic conditions that
\begin{align}\label{asymp1}
\lim_{x\rightarrow \frac{h}{r}}u_x(x,h)=0,\ \ \text{and}\ \ \lim_{x\rightarrow\frac{h}{r}}u(x,h)=-\frac{1}{r\beta}.
\end{align}
For each $h\geq 0$, we expect that the value function $x\mapsto u(x,h)$ is strictly concave for $0\leq x<\frac{h}{r}$ and the dual transform method in the previous sections can still be applied on this interval $[0,\frac{h}{r})$. In view of the set $\mathcal{C}$ when $\lambda=1$, we will now consider $y>0$ for the dual problem and define
\begin{align*}
v(y,h):=\sup_{0\leq x<\frac{h}{r}} \big( u(x,h)-xy \big),\ \ y>0,
\end{align*}

As a consequence of \eqref{asymp1}, we have the asymptotic conditions that
\begin{align}\label{yto0cond}
\lim_{y\rightarrow 0}v_y(y,h)= -\frac{h}{r}\ \ \text{and}\ \ \lim_{y \rightarrow 0} \Big(v(y,h)-yv_y(y,h)\Big)=-\frac{1}{r\beta},
\end{align}
which are different from \eqref{yhcond-1} for $0<\lambda<1$.

Based on the same analysis in the case $0<\lambda<1$, we can write down the linear dual ODE for the case $\lambda=1$ that
\begin{equation} \label{EDPtildeU3}
 \frac{\kappa^2}{2} y^2 v_{yy} - r v =\left\{
\begin{aligned}
& \frac{1}{\beta}e^{\beta h}, & & \mbox{if } y \geq e^{\beta h}, \\
& \frac{1}{\beta}y - y \left(\frac{1}{\beta} \ln y - h\right),  & & \mbox{if } 1< y < e^{\beta h}, \\
& \frac{1}{\beta} + hy,  & & \mbox{if } 0<y \leq 1,
\end{aligned}
\right.
\end{equation}

By following the same arguments to prove Proposition \ref{dual_value_function}, and replacing the free boundary condition \eqref{freedul} by the new boundary condition \eqref{yto0cond} as $y\rightarrow 0$ in the third region, we can establish the next result.

\begin{prop}
Let $h\geq 0$ be a given parameter, the ODE \eqref{EDPtildeU3} admits the unique solution explicitly by
\begin{align*}
v(y,h) =\left\{
\begin{aligned}
& C_2(h) y^{r_2} - \frac{1}{r\beta} e^{\beta h},  & & \mbox{if } y \geq e^{\beta h}, \\
& C_3(h) y^{r_1} + C_4(h) y^{r_2}-\frac{y}{r\beta}+\frac{y}{r\beta } \left(\ln y-\beta h+\frac{\kappa^2}{2r}\right),  & & \mbox{if }1 < y < e^{\beta h}, \\
& C_5(h)y^{r_1} -\frac{1}{r}hy-\frac{1}{r\beta},  & & \mbox{if } 0< y \leq 1,
\end{aligned}
\right.
\end{align*}
where $C_i(h)$, $i=2,3,4,5$ are defined in \eqref{C2h}, \eqref{C3h}, \eqref{C4h} and $\eqref{C5h}$ in Proposition \ref{dual_value_function} by setting $\lambda=1$.
\end{prop}

%
%

By using the dual value function $v(y,h)$ and applying the inverse transform that $$u(x,h)=\inf_{y>0}\Big(v(y,h)+xy\Big)$$ for $0\leq x<\frac{h}{r}$ and $u(x,h)=-\frac{1}{r\beta}$ for $x\geq\frac{h}{r}$, we can readily get the next result.

\begin{cor} \label{value_function_3}
For $(x,h)\in\mathbb{R}_+^2$ and $\lambda=1$, let us define the boundaries
\begin{align*}
\bar{x}_{\text{zero}}(h):=-e^{\beta h(r_2-1)}C_2(h)r_2,
\end{align*}
and
\begin{align*}
\bar{x}_{\text{aggr}}(h):=-C_3(h) r_1 -C_4(h) r_2+ \frac{h}{r}-\frac{\kappa^2}{2r^2\beta},
\end{align*}
and the piecewise function
\begin{align*}
& f(x,h)=\left\{
\begin{aligned}
& \Big( -x/C_2(h) r_2 \Big)^{\frac{1}{r_2-1}},  & & \mbox{if } x \leq \bar{x}_{\text{zero}}(h), \\
& \bar{f}_2(x,h),  & & \mbox{if } \bar{x}_{\text{zero}}(h) < x < \bar{x}_{\text{aggr}}(h), \\
& \Big(h-xr/C_5(h)r_1r\Big)^{\frac{1}{r_1-1}},  & & \mbox{if } \bar{x}_{\text{aggr}}(h) \leq x< \frac{h}{r},
\end{aligned}
\right.
\end{align*}
where $\bar{f}_2(x,h)$ is uniquely determined by
\begin{align*}
x=-C_3(h) r_1 (\bar{f}_2(x,h))^{r_1-1} -C_4(h) r_2 (\bar{f}_2(x,h))^{r_2-1} - \frac{1}{r\beta } \left(\ln \bar{f}_2(x,h)-\beta h +\frac{\kappa^2}{2r}\right),
\end{align*}
The value function $u(x,h)$ of the control problem in \eqref{primalvalue} can be explicitly expressed by
\begin{align*}
& u(x,h) \nonumber \\
=&\left\{
\begin{aligned}
& C_2(h) \left( \frac{-x}{C_2(h) r_2} \right)^{\frac{r_2}{r_2-1}} - \frac{1}{r\beta} e^{\beta h} + x \left( \frac{-x}{C_2(h) r_2} \right)^{\frac{1}{r_2-1}},  & & \mbox{if } x \leq \bar{x}_{\text{zero}}(h), \\
& C_3(h) (f(x,h))^{r_1} + C_4(h) (f(x,h))^{r_2} & & \\
& +\frac{f(x,h)}{r\beta } \Big( \ln f(x,h)-\beta h+\frac{\kappa^2}{2r} -1+xr\beta \Big),  & & \mbox{if } \bar{x}_{\text{zero}}(h) < x < \bar{x}_{\text{aggr}}(h), \\
& C_5(h) \left(\frac{\frac{h}{r}-x}{C_5(h)r_1}\right)^{\frac{r_1}{r_1-1}} -\frac{1}{r}h \left(\frac{\frac{h}{r}-x}{C_5(h)r_1}\right)^{\frac{1}{r_1-1}} & &\\
&-\frac{1}{r\beta}  + x \left(\frac{\frac{h}{r}-x}{C_5(h)r_1}\right)^{\frac{1}{r_1-1}},  & & \mbox{if } \bar{x}_{\text{aggr}}(h) \leq x< \frac{h}{r},\\
&-\frac{1}{r\beta},& & \mbox{if }  \frac{h}{r}\leq x.
\end{aligned}
\right.
\end{align*}
The feedback functions of the optimal consumption and portfolio are
\begin{align*}
c^{*}(x,h) =\left\{
\begin{aligned}
&0, & & \mbox{if } x \leq \bar{x}_{\text{zero}}(h), \\
&- \frac{1}{\beta}\ln f(x,h) + h,  & & \mbox{if } \bar{x}_{\text{zero}}(h) < x < \bar{x}_{\text{aggr}}(h), \\
&h,  & & \mbox{if } \bar{x}_{\text{aggr}}(h) \leq x,
\end{aligned}
\right.
\end{align*}
and
\begin{align*}
& \pi^{*}(x,h) =\frac{\mu-r}{\sigma^2} \left\{
\begin{aligned}
&(1-r_2)x, & & \mbox{if } x \leq \bar{x}_{\text{zero}}(h), \\
&\frac{2r}{\kappa^2}C_3(h)f^{r_1-1}(x,h)+\frac{2r}{\kappa^2}C_4(h)f^{r_2-1}(x,h) & & \\
&+\frac{1}{r\beta},  & & \mbox{if } \bar{x}_{\text{zero}}(h) < x < \bar{x}_{\text{aggr}}(h), \\
& \frac{2r}{\kappa^2r_1}\left(\frac{h}{r}-x\right),  & & \mbox{if } \bar{x}_{\text{aggr}}(h) \leq x< \frac{h}{r},\\
& 0,  & & \mbox{if } \frac{h}{r} \leq x,\\
\end{aligned}
\right.
\end{align*}
and the resulting consumption running maximum process is constant that $H_t^*=H_0^*=h$ for $t>0$.
\end{cor}

\section{Proof of Lemma \ref{Lipschitz_c_pi}} \label{appB} 
\proof By \eqref{feedbackcp}, \eqref{feedbackpi} and the inverse transform, we can express $c^*$ and $\pi^*$ in terms of the primal variables as in \eqref{consumption:primal} and \eqref{invest:primal}. Combining the expressions of $c^*$ and $\pi^*$ with Proposition \ref{dual_value_function} (implying that the coefficient functions $(C_i(h))_{2 \leq i \leq 5}$ are $C^1$),  Lemma \ref{f_regularity} (implying the $C^1$ regularity of $f$), together with the continuity of $f$ at the boundary between the three regions, we can draw the conclusion that $(x,h) \mapsto c^*(x,h)$ and $(x,h) \mapsto \pi^*(x,h)$ are locally Lipschitz on $\mathcal{C}$.

Now in order to prove the Lipschitz property of $\pi^*$, we show separately that the partial derivatives $\frac{\partial \pi^*}{\partial x}$ and $\frac{\partial \pi^*}{\partial h}$ are bounded. 
\ \\
\ \\
\textit{Step 1}: the boundedness of  $\frac{\partial \pi^*}{\partial x}$.

First, using $\pi^*$ in \eqref{invest:primal}, we have 
\begin{align}\label{invest_derivative:primal}
& \frac{\partial \pi^*}{\partial x}(x,h) =\frac{\mu-r}{\sigma^2}\nonumber \\
&\times\left\{
\begin{aligned}
&(1-r_2), & & \mbox{if } x \leq x_{\text{zero}}(h), \\
&\frac{2r}{\kappa^2}C_3(h) (r_1-1) f_2^{r_1-2}(x,h) \frac{\partial f_2}{\partial x} & & \\
&~~~~~~~~~~~~~~~ +\frac{2r}{\kappa^2}C_4(h)(r_2-1)f_2^{r_2-2}(x,h) \frac{\partial f_2}{\partial x},  & & \mbox{if } x_{\text{zero}}(h) < x < x_{\text{aggr}}(h), \\
& \frac{2r}{\kappa^2}C_5(h) (r_1-1) f_3^{r_1-2}(x,h) \frac{\partial f_3}{\partial x} & & \\
&~~~~~~~~~~~~~~~+\frac{2r}{\kappa^2}C_6(h)(r_2-1)f_3^{r_2-2}(x,h) \frac{\partial f_3}{\partial x},  & & \mbox{if } x_{\text{aggr}}(h) \leq x \leq x_{\text{lavs}}(h).
\end{aligned}
\right.
\end{align}
Note that the first line is constant and hence bounded. For the second line, by differentiating \eqref{f2-equt} and using $r_1(r_1-1)=r_2(r_2-1)=\frac{2r}{\kappa^2}$, we have that 
$$
1= - \frac{2r}{\kappa^2} C_3(h) f_2^{r_1-2}(x,h) \frac{\partial f_2}{\partial x} - \frac{2r}{\kappa^2} C_4(h) f_2^{r_2-2}(x,h) \frac{\partial f_2}{\partial x} - \frac{1}{r \beta} \frac{1}{f_2} \frac{\partial f_2}{\partial x}.
$$
Plugging this back to $\frac{\partial \pi^*}{\partial x}$, we can obtain
\begin{align*}
\frac{\partial \pi^*}{\partial x}(x,h) &= \frac{\mu-r}{\sigma^2} \Big( \frac{2r}{\kappa^2} C_3(h) f_2^{r_1-2}(x,h) \frac{\partial f_2}{\partial x} (r_1-r_2) + (1-r_2) + (1-r_2) \frac{1}{r \beta} \frac{1}{f_2} \frac{\partial f_2}{\partial x} \Big) \\
&= \frac{\mu-r}{\sigma^2} \Big(  \big( \frac{2r}{\kappa^2} C_3(h) f_2^{r_1-1}(x,h) (r_1-r_2) + (1-r_2) \frac{1}{r \beta} \big) \frac{1}{f_2} \frac{\partial f_2}{\partial x} + (1-r_2) \Big).
\end{align*}
Combining the above with \eqref{f_regularity}, we can obtain 
\begin{equation} \label{eq:invest_derivative_1}
\frac{\partial \pi^*}{\partial x}(x,h) =  \frac{\mu-r}{\sigma^2} \Big( \frac{A}{B} + (1-r_2) \Big),
\end{equation}
where 
\begin{align} \label{eq:def_A_B}
& A:=  \frac{2r}{\kappa^2} C_3(h) f_2^{r_1-1}(x,h) (r_1-r_2) + \frac{1-r_2}{r \beta}, \nonumber \\
& B:=  \frac{2r}{\kappa^2} \Big( -C_3(h) f_2^{r_1-1}(x,h) - C_4(h) f_2^{r_2-1}(x,h) \Big) - \frac{1}{r \beta}.
\end{align}
In what follows, we show that there exist two constants $A_0 > 0, B_0 < 0$, independent from $h$, such that $0 \leq A \leq A_0$ and $B \leq B_0$. Combining with \eqref{eq:invest_derivative_1}, this shows that 
$$
\frac{\mu-r}{\sigma^2} \left( \frac{A_0}{B_0} + (1-r_2) \right) \leq \frac{\partial \pi^*}{\partial x}(x,h) \leq \frac{\mu-r}{\sigma^2} (1-r_2),
$$
hence $\frac{\partial \pi^*}{\partial x}$ is bounded.

Indeed, as $C_3< 0$ and $r_1 > 1 > r_2$, the map $y \mapsto \frac{2r}{\kappa^2} C_3(h) y^{r_1-1} (r_1-r_2) + (1-r_2) \frac{1}{r \beta}$ is decreasing. Plugging the lower and upper bounds of $f_2$, $e^{(\lambda-1) \beta h}$ and $e^{\lambda \beta h}$, we have that
$$
A \geq \frac{2r}{\kappa^2} C_3(h) e^{\lambda(r_1-1) \beta h} (r_1-r_2) + (1-r_2) \frac{1}{r \beta}=0,
$$
and 
\begin{align*}
A &\leq  \frac{2r}{\kappa^2} C_3(h) e^{\lambda(r_1-1) \beta h} (r_1-r_2) + (1-r_2) \frac{1}{r \beta} \\
& = (1-r_2) \frac{1}{r \beta} \left( 1- e^{-(r_1-1) \beta h} \right) \leq (1-r_2) \frac{1}{r \beta}:=A_0
\end{align*}
For $B$, from the proof of Lemma \ref{vyy_positive}, we know that  in the closed interval $[e^{(\lambda-1) \beta h}, e^{\lambda \beta h} ]$, the map $z: y \mapsto  \frac{2r}{\kappa^2} ( -C_3(h) y^{r_1-1} - C_4(h) y^{r_2-1} ) - \frac{1}{r \beta}$ is strictly negative. Moreover, it is either monotone or first decreasing then increasing, as shown in \textit{Step 2} of the proof of Lemma \ref{vyy_positive}. Therefore, the upper bound of $z$ is attained at either $y= e^{\lambda \beta h}$ or $y= e^{(\lambda-1) \beta h}$. 

At the boundary $y=e^{\lambda \beta h}$, we have 
\begin{align*}
z(e^{\lambda \beta h}) & = - \frac{(1-\lambda)^{r_1-r_2}}{(r_1-r_2) \beta r} \frac{2r}{\kappa^2} \quad\times \Bigg( \frac{1}{1-r_2} e^{-(1-r_2)\beta h} -\frac{\lambda}{\lambda(1-r_2)-(r_1-r_2)} e^{ -(r_1-r_2) \beta h} \Bigg) \\
&~~~~~~~~~~~~~~~~~~~~~-\frac{1}{(r_1-r_2) \beta r} (r_1-1) \left( 1 - e^{-(1-r_2) \beta h} \right) : = \mathcal{L}_1(h).
\end{align*}
The above is strictly negative for $h \in [H_0, +\infty)$. To wit, when $h \rightarrow +\infty$, we have that the limit
$\lim_{h \rightarrow +\infty}  \mathcal{L}_1(h) = -\frac{r_1-1}{r_1-r_2} \frac{1}{\beta r} < 0$.
Moreover, we have 
$$
 \mathcal{L}_1(0) = - \frac{(1-\lambda)^{r_1-r_2}}{(r_1-r_2) \beta r} \frac{2r}{\kappa^2} \quad\times \Bigg( \frac{1}{1-r_2} -\frac{\lambda}{\lambda(1-r_2)-(r_1-r_2)} \Bigg) < 0.
$$
We now analyze the monotonicity of each term in the function $h \mapsto \mathcal{L}_1(h)$. The first term in $\mathcal{L}_1$ is strictly increasing on $[0, + \infty)$ and takes values in the interval $[ \mathcal{L}_1(0),0)$. The second term in $\mathcal{L}_1$ is strictly decreasing on $[0, + \infty)$ with values in the interval $(\lim_{h \rightarrow +\infty}  \mathcal{L}_1(h),0]$. Hence we can define 
$
B_0':= \sup_{H_0 \leq h < +\infty}  \mathcal{L}_1(h) < 0.$ At $y=e^{(\lambda-1) \beta h}$, we have 
\begin{align*}
z(e^{(\lambda-1) \beta h}) & = - \frac{(1-\lambda)^{r_1-r_2}}{(r_1-r_2) \beta r} \frac{2r}{\kappa^2} \quad\times \Bigg( \frac{1}{1-r_2} -\frac{\lambda}{\lambda(1-r_2)-(r_1-r_2)} e^{ -(r_1-1) \beta h} \Bigg) \\
&~~~~~~~~~~~~~~~~~~~~~-\frac{1}{(r_1-r_2) \beta r} (1-r_2) \left( 1 - e^{-(r_1-1) \beta h} \right) : = \mathcal{L}_2(h).
\end{align*}
Similar arguments as the point $e^{\lambda \beta h}$ lead to 
$B_0'':= \sup_{H_0 \leq h < +\infty}  \mathcal{L}_2(h) < 0$. We can finally set  $B_0: = \max \{ B_0', B_0'' \}$

For the third line of \eqref{invest_derivative:primal}, similar calculations as for the second line lead to
\begin{align*}
\frac{\partial \pi^*}{\partial x}(x,h) &=  \frac{\mu-r}{\sigma^2} \Big(  \big( \frac{2r}{\kappa^2} C_5(h) f_3^{r_1-1}(x,h) (r_1-r_2)  \big) \frac{1}{f_3} \frac{\partial f_3}{\partial x} + (1-r_2) \Big) \\
&=  \frac{\mu-r}{\sigma^2} \Big( \frac{A'}{B'} + (1-r_2) \Big),
\end{align*}
where 
\begin{equation} \label{eq:def_A_B_prime}
A':=\frac{2r}{\kappa^2} C_5(h) f_3^{r_1-1}(x,h) (r_1-r_2), B':=  \frac{2r}{\kappa^2} \Big( -C_5(h) \big(f_3(x,h)\big)^{r_1-1} - C_6(h) \big(f_3(x,h)\big)^{r_2-1} \Big).
\end{equation}
 We claim that there exist two constants $A_1 > 0, B_1 < 0$, which are independent from $h$ and satisfy that $0 \leq A' \leq A_1$ and $B' \leq B_1$. This implies that 
$$
\frac{\mu-r}{\sigma^2} \left( \frac{A_1}{B_1} + (1-r_2) \right) \leq \frac{\partial \pi^*}{\partial x}(x,h) \leq \frac{\mu-r}{\sigma^2} (1-r_2),
$$
hence $\frac{\partial \pi^*}{\partial x}$ is bounded in the third region.

Now we find two constants $A_1$ and $B_1$. As $C_5 > 0$ and $r_1 > 1 > r_2$, we know that the map $y \mapsto \frac{2r}{\kappa^2} C_5(h) y^{r_1-1} (r_1-r_2)$ is increasing and clearly $A' \geq 0$. Using the condition that $(1-\lambda) e^{(\lambda-1) \beta h} \leq f_3 \leq e^{(\lambda-1) \beta h}$, we have
\begin{align*}
A'  &\leq  \frac{2r}{\kappa^2} C_5(h) \left( e^{(\lambda-1) \beta h} \right)^{r_1-1} (r_1-r_2) \\
&=\frac{1-r_2}{(r_1-r_2) \beta r} \left( 1 - e^{(1-r_1) \beta h} \right) \leq \frac{1-r_2}{(r_1-r_2) \beta r}:= A_1 > 0.
\end{align*}

As $C_5 > 0, C_6 > 0$ and $(1-\lambda) e^{(\lambda-1) \beta h} \leq f_3 \leq e^{(\lambda-1) \beta h}$, we have that 
\begin{align*}
B'  \leq & -\frac{2r}{\kappa^2} C_5(h) \left( (1-\lambda) e^{(\lambda-1)\beta h} \right)^{r_1-1} - \frac{2r}{\kappa^2} C_6(h) \left( e^{(\lambda-1)\beta h} \right)^{r_2-1}  \\
=&   - (1-\lambda)^{r_1-1} \frac{1-r_2}{(r_1-r_2) \beta r} \left( 1 - e^{(1-r_1) \beta h} \right) \\
&- \frac{2r}{\kappa^2} \frac{(1-\lambda)^{r_1-r_2}}{(r_1-r_2) \beta r} \left( \frac{1}{1-r_2} - \frac{\lambda}{\lambda(1-r_2) - (r_1-r_2)} e^{(1-r_1) \beta h} \right): =  \mathcal{L}_3(h).
\end{align*}
As $h \mapsto \mathcal{L}_3(h)$ is strictly decreasing, we have 
$$
B' \leq \mathcal{L}_3(0) = - \frac{2r}{\kappa^2} \frac{(1-\lambda)^{r_1-r_2}}{(r_1-r_2) \beta r} \left( \frac{1}{1-r_2} - \frac{\lambda}{\lambda(1-r_2) - (r_1-r_2)} \right): = B_1 < 0.
$$

\ \\
\textit{Step 2}: the boundedness of $\frac{\partial \pi^*}{\partial h}$.

First, using equation \eqref{f_derivative_z} and the definition of $g$ in equation \eqref{sol_for_x}, we have
\begin{align}
& f_h(x,h) = -g_h(f,h) \cdot f_x(x,h) \nonumber \\
=&\left\{
\begin{aligned}
&  -C_2'(h) r_2 f_1(x,h)^{r_2-1} \cdot \frac{1}{-C_2(h)\frac{2r}{\kappa^2} f_1(x,h)^{r_2-2}},  & & \mbox{if } x \leq x_{\text{zero}}(h), \\
& \left( C_3'(h) r_1 f_2(x,h)^{r_1-1} + C_4'(h) r_2 f_2(x,h)^{r_2-1} - \frac{\lambda \beta}{r \beta} \right) \cdot & & \\
& \frac{1}{-C_3(h) \frac{2r}{\kappa^2} f_2(x,h)^{r_1-2} - C_4(h) \frac{2r}{\kappa^2} f_2(x,h)^{r_2-2} - \frac{1}{r \beta f_2(x,h)}},  & & \mbox{if } x_{\text{zero}}(h) < x < x_{\text{aggr}}(h), \\
& \left( C_5'(h) r_1 f_3(x,h)^{r_1-1} + C_6'(h) r_2 f_3(x,h)^{r_2-1} - \frac{1}{r} \right) \cdot & &\\
&\frac{1}{-C_5(h) \frac{2r}{\kappa^2} f_3(x,h)^{r_1-2} - C_6(h) \frac{2r}{\kappa^2} f_3(x,h)^{r_2-2} },  & & \mbox{if } x_{\text{aggr}}(h) \leq x \leq x_{\text{lavs}}(h).\\
\end{aligned} 
\right.  \nonumber
\end{align}

We analyze the derivative $\frac{\partial \pi^*}{\partial h}$ in different regions separately. In the region $x \leq x_{\text{zero}}(h)$, $\frac{\partial \pi^*}{\partial h}=0$, hence is bounded. 

In the region $x_{\text{zero}}(h) < x < x_{\text{aggr}}(h)$, we have
\begin{align*}
\frac{\partial \pi^*}{\partial h}  = \ & \frac{\mu-r}{\sigma^2} \Big( \frac{2r}{\kappa^2} C_3'(h) f_2(x,h)^{r_1-1} + \frac{2r}{\kappa^2} C_3(h) (r_1-1) f_2(x,h)^{r_1-2} \frac{\partial f_2}{\partial h} \\
 &  + \frac{2r}{\kappa^2} C_4'(h) f_2(x,h)^{r_2-1} + \frac{2r}{\kappa^2} C_4(h) (r_2-1) f_2(x,h)^{r_2-2} \frac{\partial f_2}{\partial h}  \Big).
\end{align*}

By differentiating \eqref{f2-equt} and using $r_1(r_1-1)=r_2(r_2-1)=\frac{2r}{\kappa^2}$, we have that 
\begin{align*}
& C_4'(h) \frac{2r}{\kappa^2} f_2(x,h)^{r_2-1} + C_4(h) \frac{2r}{\kappa^2} (r_2-1) \frac{\partial f_2}{\partial h}  f_2(x,h)^{r_2-2}   \\
= \ &  - C_3'(h) r_1 (r_2-1)  f_2(x,h)^{r_1-1} - C_3(h) \frac{2r}{\kappa^2} (r_2-1) \frac{\partial f_2}{\partial h}  f_2(x,h)^{r_1-2}  \\
& ~~~~~~~~~ - \frac{1}{r \beta} (r_2-1) \left( \frac{1}{f_2} \frac{\partial f_2(x,h)}{\partial h} - \lambda \beta \right).
\end{align*}
Replacing this back to the previous expression of $\frac{\partial \pi^*}{\partial h}$, we can obtain that
\begin{align} \label{eq:h_derivative_second_zone}
\frac{\partial \pi^*}{\partial h} & = \ \frac{\mu-r}{\sigma^2} \Bigg( \Big( \frac{2r}{\kappa^2} (r_1-r_2) C_3(h) f_2(x,h)^{r_1-1} + \frac{1}{r \beta} (1-r_2) \Big) \frac{1}{f_2} \frac{\partial f_2}{\partial h} \nonumber  \\
 &  + r_1 (r_1-r_2) C_3'(h) f_2(x,h)^{r_1-1} - \frac{\lambda \beta}{r \beta} (1-r_2)  \Bigg) \nonumber \\
 & = \ \frac{\mu-r}{\sigma^2} \Big( A \cdot \frac{1}{f_2} \frac{\partial f_2}{\partial h} + r_1 (r_1-r_2) C_3'(h) f_2(x,h)^{r_1-1} - \frac{\lambda \beta}{r \beta} (1-r_2)  \Big),
\end{align}
where $A$ is defined in \eqref{eq:def_A_B}. In \eqref{eq:h_derivative_second_zone}, the third term is a constant. For the second term, we can calculate that $C_3'(h)$ and $C_4'(h)$ are given respectively by 
$$
C_3'(h) = \frac{1}{(r_1-r_2) \beta r} \lambda \beta e^{\lambda (1-r_1) \beta h} > 0,
$$
and
\begin{align*}
C_4'(h) &= \frac{(1-\lambda)^{r_1-r_2}}{(r_1-r_2) \beta r} \left( (\lambda-1) \beta e^{(\lambda-1) (1-r_2) \beta h} - \lambda \beta e^{( \lambda (1-r_2) - (r_1-r_2) ) \beta h} \right) \\
& - \frac{1}{(r_1-r_2) \beta r} (\lambda-1) \beta e^{(\lambda-1) (1-r_2) \beta h}.
\end{align*}
Using the fact that $e^{(\lambda-1) \beta h} < f_2(x,h) < e^{\lambda \beta h}$ and $r_1 > 1$, we have
$$
0 \leq \frac{\lambda r_1}{r} e^{-\beta h (r_1-1)} \leq r_1 (r_1-r_2) C_3'(h) f_2(x,h)^{r_1-1} \leq \frac{\lambda r_1}{r},
$$
hence the second term is also bounded. Now we consider the first term in the bracket of \eqref{eq:h_derivative_second_zone}. From the previous calculations, we know that $0 \leq A \leq A_0$, hence it is enough to show that $\frac{1}{f_2} \frac{\partial f_2}{\partial h}$ is bounded. Indeed, we have
$$
\frac{1}{f_2} \frac{\partial f_2}{\partial h} = \left( C_3'(h) r_1 f_2(x,h)^{r_1-1} + C_4'(h) r_2 f_2(x,h)^{r_2-1} - \frac{\lambda \beta}{r \beta} \right) \cdot \frac{1}{B},
$$
where $B$ is defined in \eqref{eq:def_A_B}. From the expression of $C_4'(h)$, we can derive that 
\begin{align*}
C_4'(h) r_2 f_2(x,h)^{r_2-1} & \geq r_2 \frac{1-\lambda}{(r_1-r_2)r} \Big( 1- (1-\lambda)^{r_1-r_2} \Big) - r_2  \frac{(1-\lambda)^{r_1-r_2}}{(r_1-r_2)r} \lambda e^{-(r_1-r_2) \beta h} \\
& \geq r_2 \frac{1-\lambda}{(r_1-r_2)r} \Big( 1- (1-\lambda)^{r_1-r_2} \Big), 
\end{align*}
and
\begin{align*}
& \ C_4'(h) r_2 f_2(x,h)^{r_2-1}  \\
\leq & \  r_2 \frac{1-\lambda}{(r_1-r_2)r} \Big( 1- (1-\lambda)^{r_1-r_2} \Big) e^{-(1-r_2) \beta h} - r_2  \frac{(1-\lambda)^{r_1-r_2}}{(r_1-r_2)r} \lambda e^{-(r_1-1) \beta h}   \\
\leq & \ - r_2  \frac{(1-\lambda)^{r_1-r_2}}{(r_1-r_2)r} \lambda.
\end{align*}
Using the fact that $B \leq B_0 < 0$, we can draw the conclusion that $\frac{1}{f_2} \frac{\partial f_2}{\partial h}$ is bounded.

In the region $x_{\text{aggr}}(h) \leq x \leq x_{\text{lavs}}(h)$, similar calculations lead to
\begin{align*} 
\frac{\partial \pi^*}{\partial h} & = \ \frac{\mu-r}{\sigma^2} \Bigg( \frac{A'}{B'} \left( C_5'(h) r_1 f_3(x,h)^{r_1-1} + C_6'(h) r_2 f_3(x,h)^{r_2-1} - \frac{1}{r} \right) \nonumber  \\
 &  + r_1 (r_1-r_2) C_5'(h) f_3(x,h)^{r_1-1} - \frac{1}{r} (1-r_2)  \Bigg),
\end{align*}
where $A'$ and $B'$ are defined in \eqref{eq:def_A_B_prime}. Now combining the fact that $C_5'(h)$ has terms of $e^{(\lambda-1)(1-r_1)\beta h}$ and $e^{\lambda (1-r_1) \beta h}$, $C_6'(h)$ has terms of $e^{(\lambda-1)(1-r_2)\beta h}$ and $e^{( \lambda (1-r_2) - (r_1-r_2) ) \beta h}$, $0 \leq A' \leq A_1$, $B' \leq B_1 <0$ and $(1-\lambda) e^{(\lambda-1) \beta h} \leq f_3 \leq e^{(\lambda-1) \beta h}$, we can obtain the desired boundedness result. The calculations are similar as before and we omit the details here. Putting all the pieces together, we obtain the desired boundedness of $\frac{\partial \pi^*}{\partial h}$.
\smartqed

\begin{acknowledgements}
We thank two anonymous referees for their helpful comments on the presentation of this paper. H. Pham and X. Yu appreciate the financial support by the PROCORE-France/Hong Kong Joint Research Scheme under no. F-PolyU501/17. X. Yu is partially supported by the Hong Kong Early Career Scheme under grant no. 25302116. X. Li is partially supported by the Hong Kong General Research Fund under grant no. 15213218 and no. 15215319.
\end{acknowledgements}

%
\section*{Conflict of interest}
The authors declare that they have no conflict of interest.


\begin{thebibliography}{}
%
%

\bibitem{BAY}
Angoshtari, B., Bayraktar, E., Young, V.: Optimal dividend distribution under drawdown and ratcheting constraints on dividend rates. SIAM Journal on Financial Mathematics. \textbf{10}(2), 547-577 (2019)

\bibitem{Arun}
Arun, T.: The Merton problem with a drawdown constraint on consumption. Preprint, arXiv:1210.5205 (2012)

\bibitem{Bilsen17}
Bilsen, S., Laeven, R., Nijman, T.: Consumption and portfolio choice under loss aversion and endogenous updating of the reference level. Management Science. \textbf{66}(9), 3927-3955 (2020)

\bibitem{constantinides1990habit}
Constantinides, G. M.: Habit formation: A resolution of the equity premium puzzle. Journal of Political Economy. \textbf{98}(3), 519--543 (1990)

\bibitem{CoxHuang}
Cox, J. C., Huang, C.: Optimal consumption and portfolio policies when asset prices follow a diffusion process. Journal of Economic Theory. \textbf{49}, 33--83 (1989)

\bibitem{Curatola17}
Curatola, G.: Optimal portfolio choice with loss aversion over consumption. The Quarterly Review of Economics and Finance. \textbf{66}, 345-358 (2017)

\bibitem{DepKart}
Detemple, J., Karatzas, I.: Non-addictive habits: optimal consumption-portfolio policies. Journal of Economic Theory. \textbf{113}, 265-285 (2003)

\bibitem{detemple1992optimal}
Detemple, J., Zapatero, F.: Optimal consumption-portfolio policies with habit formation. Mathematical Finance. \textbf{2}(4), 251--274 (1992)

\bibitem{Dyb}
Dybvig, P. H.: Optimal consumption and portfolio policies when asset prices follow a diffusion process. The Review of Economics Studies. \textbf{62}(2), 287--313 (1995)

\bibitem{ElieTouzi}
Elie, R., Touzi, N.: Optimal lifetime consumption and investment under a drawdown constraint. Finance and Stochastics. \textbf{12}, 299-330 (2008)

\bibitem{englezos2009utility}
Englezos, N., Karatzas, I.: Utility maximization with habit formation: Dynamic programming and stochastic PDEs. SIAM Journal on Control and Optimization. \textbf{48}(2), 481--520 (2009)

\bibitem{GHR}
Guasoni, P., Huberman, G., Ren, D.: Shortfall aversion. Mathematical Finance. \textbf{30}(3), 869--920 (2020)

\bibitem{He3}
He, X., Strub, M.: How endogenization of the reference point affects loss aversion: a study of portfolio selection. Preprint, available at SSRN:https://dx.doi.org/10.2139/ssrn.3318295 (2019)

\bibitem{He2}
He, X., Yang, L.: Realization utility with adaptive reference points. Mathematical Finance. \textbf{29}(2), 409-447 (2019)

\bibitem{He1}
He, X., Zhou, X. Y.: Myopic loss aversion, reference point, and money illusion. Quantitative Finance. \textbf{14}(9), 1541-1554 (2014)

\bibitem{LiuH}
Liu, H.: Optimal consumption and investment with transaction costs and multiple risky assets. The Journal of Finance. \textbf{59}(1), 289-338 (2004)

\bibitem{Meh}
Mehra, R., Prescott, E. C.: The equity premium: A puzzle. Journal of Monetary Economics. \textbf{15}(2), 145-161 (1969).

\bibitem{Mert1}
Merton, R. C.: Lifetime portfolio selection under uncertainty: the continuous time case. The Review of Economics and Statistics. \textbf{51}(3), 247-257 (1969)

\bibitem{Mert2}
Merton, R. C.: Optimal consumption and portfolio rules in a continuous-time model. Journal of Economic Theory. \textbf{3}, 373-413 (1971)

\bibitem{munk2008portfolio}
Munk, C.: Portfolio and consumption choice with stochastic investment opportunities and habit formation in preferences. Journal of Economic Dynamics and Control. \textbf{32}(11), 3560--3589 (2008)

\bibitem{schroder2002isomorphism}
Schroder, M., Skiadas, C.: An isomorphism between asset pricing models with and without linear habit formation. The Review of Financial Studies. \textbf{15}(4), 1189--1221 (2002)

\bibitem{tvekah92}
Tversky, A., Kahneman, D.: Advances in prospect theory: Cumulative representation of uncertainty. Journal of Risk and uncertainty. \textbf{5}(4), 297--323 (1992)

\bibitem{Vayanos}
Vayanos, D.: Transaction costs and asset prices: a dynamic equilibrium model. The Review of Financial Studies. \textbf{11}(1), 1-58 (1998)
%
%
\bibitem{yu2015utility}
Yu, X.: Utility maximization with addictive consumption habit formation in incomplete semimartingale markets. The Annals of Applied Probability. \textbf{25}(3), 1383--1419 (2015)

\bibitem{yu2017}
Yu, X.: Optimal consumption under habit formation in markets with transaction costs and random endowments. The Annals of Applied Probability. \textbf{27}(2), 960-1002 (2017)
\end{thebibliography}


\end{document}